\documentclass[aps,prd,10pt,twocolumn,nofootinbib,superscriptaddress]{revtex4-2}

\usepackage[T1]{fontenc}
\usepackage{lmodern}
\usepackage[utf8]{inputenc}
\usepackage{amsmath,amssymb}
\usepackage{bm}
\usepackage{graphicx}
\usepackage[hidelinks]{hyperref}

\newcommand{\epsnn}{\epsilon_{nn'}}
\newcommand{\dm}{\delta m}
\newcommand{\mun}{\mu_n}

\newcommand{\Non}{N_{\mathrm{on}}}
\newcommand{\Noff}{N_{\mathrm{off}}}
\newcommand{\Slock}{S_{\mathrm{lock}}}

\newcommand{\dd}{\mathrm{d}}

\begin{document}

	\title{Sensitivity and search strategies for neutron–hidden-neutron oscillations with very-cold-neutron interferometry}

	\author{A. Capolupo}
	\email{capolupo@sa.infn.it}
	\affiliation{Dipartimento di Fisica ``E.R. Caianiello'', Università di Salerno,
		and INFN -- Gruppo Collegato di Salerno, Via Giovanni Paolo II, 132,
		84084 Fisciano (SA), Italy}

	\author{P. B\"{o}ni }
	\email{peter.boeni@tum.de}
	\affiliation{Technical University of Munich, TUM School of Natural Sciences, Physics Department, 85748 Garching, Germany}
	\author{G. Pisacane}
	\email{gpisacane@unisa.it}
	\affiliation{Dipartimento di Fisica ``E.R. Caianiello'', Università di Salerno,
		and INFN -- Gruppo Collegato di Salerno, Via Giovanni Paolo II, 132,
		84084 Fisciano (SA), Italy}

\begin{abstract}
We investigate the sensitivity of very-cold-neutron Mach--Zehnder
interferometry to neutron--hidden-neutron oscillations when the mass
splitting is unknown. We connect the resonant disappearance signal
to the statistics of the detected neutrons and compare paired,
three-field, and block acquisition protocols at fixed scheduled beam
time. The comparison incorporates a common model of temporal rate
drift, field-dependent transmission, and counting fluctuations,
together with the time required to change and stabilize the magnetic
field. We determine the projected extension of existing laboratory
exclusion limits and examine how the accepted neutron phase space
and the calibration of the two interferometer outputs affect the
accessible mixing strength.
\end{abstract}

\maketitle

\section{Introduction}

The particle nature of dark matter remains an open
question~\cite{DS5}. A class of extensions of the Standard
Model introduces a mirror sector, whose baryonic states can
contribute to the dark matter abundance~\cite{MM1,MM2,MM4,MM3,MM5}.
More generally, hidden-sector extensions may contain a neutral
baryonic state $n'$ coupled to the ordinary neutron $n$ through
an off-diagonal mixing term. The propagation eigenstates are
then mixtures of the two states, allowing an initially ordinary
neutron to convert into the hidden
state~\cite{BerezhianiBento2006,BerezhianiReview}.
Laboratory searches exploit the resulting disappearance of
ordinary neutrons or their regeneration after propagation
through an absorber. These include ultracold-neutron (UCN)
storage measurements and their phenomenological
interpretations~\cite{Nesvizhevsky2006,Ban2007,Serebrov2009,BerezhianiNesti2012,Abel2021,AyresMass2026},
UCN beam experiments~\cite{Ban2023,Brenot2026}, and reactor-
or spallation-based
searches~\cite{Sarrazin2015,Sarrazin2016,StasserMURMUR2021,AlmazanSTEREO2022,Broussard2022,GonzalezSNS2024}.
The phenomenology of hidden baryons also connects regeneration
searches to nuclear absorption processes~\cite{MM6}.
The sensitivity of each method depends on the neutron energy,
propagation time, magnetic field, and experimental geometry.
The searches are commonly compared in the plane
$(\dm,\tau_{nn'})$, where $\dm=m_{n'}-m_n$ is the vacuum mass
splitting between the hidden and ordinary neutron, and
$\tau_{nn'}=\hbar/\epsnn$ is the characteristic oscillation
time associated with the real mixing energy $\epsnn$,
taken to be nonnegative by a choice of relative phase.
We express mass splittings in energy units ($c=1$ in the
Hamiltonian) and retain the reduced Planck constant $\hbar$
explicitly in the time evolution; in the nonrelativistic
kinematic and optical-potential formulas, $m_n$ denotes the
neutron inertial mass in consistent units.

Interferometers based on multilayer mirrors and diffraction
gratings provide complementary approaches to neutron phase
measurements, including operation in the very-cold-neutron
regime~\cite{Ebisawa,Zeilinger,Fujiie,Ackermann2026}.
The very-cold-neutron (VCN) Mach--Zehnder proposal of
Ref.~\cite{OriginalProposal} exploits the long interaction
time of slow neutrons and the possibility of controlling
their energy through an applied magnetic field.
For a selected spin projection, the ordinary and hidden
neutron energies become degenerate when the Zeeman shift
compensates $\dm$. The resulting enhancement of conversion
into the hidden state produces a deficit in the detected
ordinary-neutron intensity. Comparing the intensities at
the resonance and at a nearby reference field provides a
means of identifying this deficit, while the two
interferometer outputs also give access to changes in the
relative phase of the ordinary-neutron amplitudes.

The two-port geometry thus combines a resonant disappearance
measurement with a phase-sensitive observable. In a
disappearance search, a reduction of the detected rate must
be distinguished from a variation of the incident flux or
the apparatus response. Moreover, when the mass splitting
is unknown, the available beam time must be distributed
among different magnetic configurations. The sensitivity
of the search therefore depends not only on the resonant
conversion probability but also on the counting statistics,
the correlations of instrumental uncertainties, and the
acquisition protocol.

Starting from the interferometric proposal of
Ref.~\cite{OriginalProposal}, we develop a statistical
description of the conversion signal and determine the
sensitivity of a VCN search at fixed total beam time.
We use the complete interferometer response for the
fixed-calibration calculation and a reduced propagation
model to isolate the effects of effective magnetic-field
broadening and relative calibration. We first determine
the sensitivity at a given mass splitting and extend the
local projections over a broader region of parameter space.
We then consider a magnetic-field scan over an unknown mass
splitting, comparing paired, symmetric three-field, and
block acquisitions within a common covariance model of
temporal rate drift and field-dependent transmission.
By accounting for the time required to change and stabilize
the magnetic field, we investigate the optimal distribution
of the available exposure and the extent to which such a
search can probe mixing beyond existing laboratory
constraints. Finally, we relate the detected neutron rate
to the accepted beam phase space and quantify the calibration
information provided by the two interferometer outputs.

The paper is organized as follows. Sections~\ref{sec:model}--\ref{sec:covariance} establish the
connection between the conversion dynamics, the instrumental
response, and the statistical inference. We determine the
local sensitivity in Sec.~\ref{sec:updated_sensitivity},
investigate the magnetic-field scanning strategies in
Sec.~\ref{sec:scan}, and examine the neutron acceptance
and the calibration of the interferometer outputs in
Sec.~\ref{sec:high_rate}.

\section{Signal model and magnetic-lock-in observable}
	\label{sec:model}

\begin{figure}[t]
		\centering
		\IfFileExists{schema_V5.png}{%
			\includegraphics[width=\columnwidth]{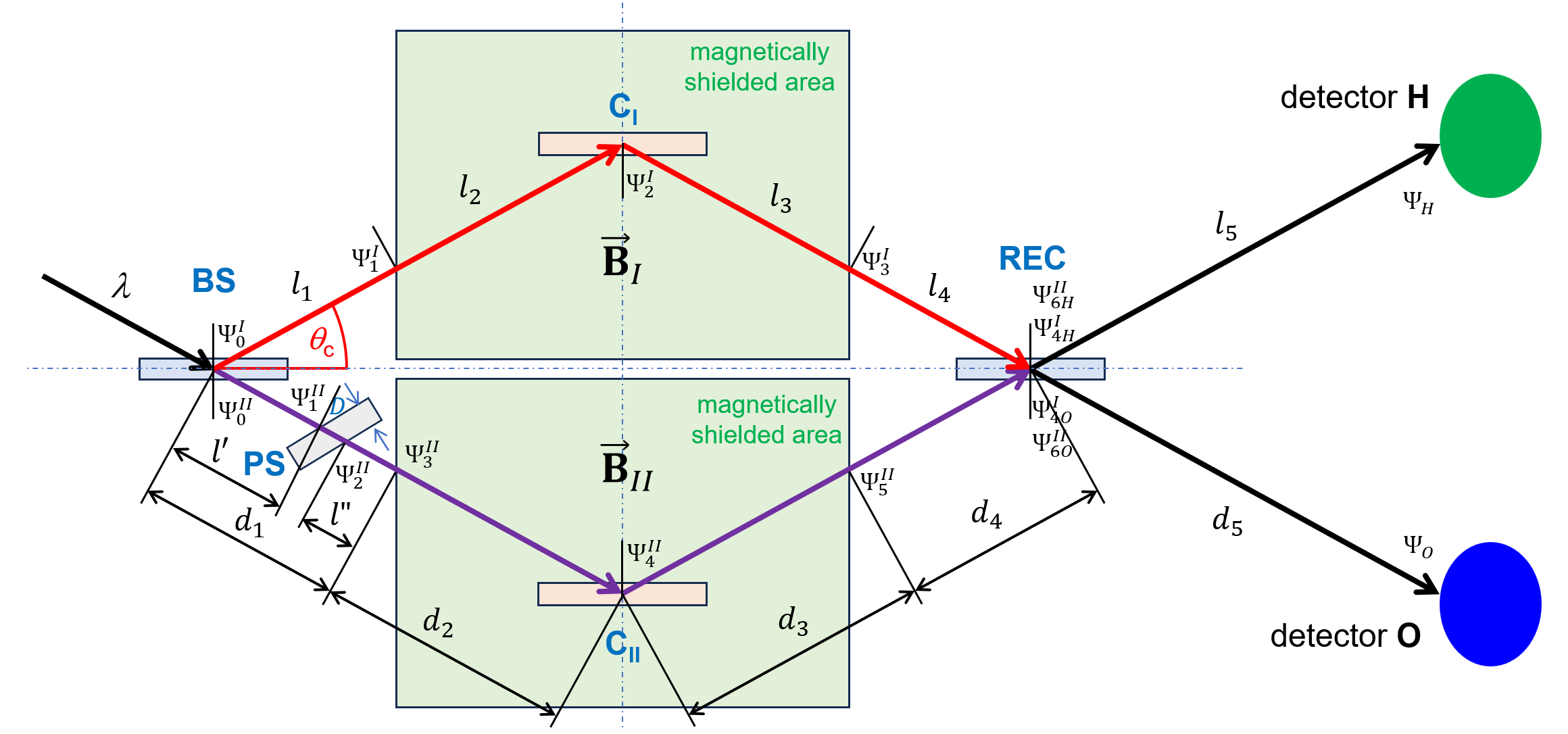}%
		}{%
			\IfFileExists{schema_V5(1).png}{%
                \includegraphics[width=\columnwidth]{schema_V5(1).png}%
            }{%
                \includegraphics[width=\columnwidth]{schema_V5(2).png}%
            }%
		}
		\caption{Schematic layout of the proposed VCN interferometer. The polarized beam is split into paths $I$ and $II$, recombined at REC, and measured at the output ports $O$ and $H$. The active magnetic fields define the tunable conversion regions; the green regions denote magnetic shielding. Reproduced from Ref.~\cite{OriginalProposal}.}
		\label{fig:schema_v5}
	\end{figure}

	The sensitivity study is based on the VCN Mach--Zehnder configuration introduced in Ref.~\cite{OriginalProposal}. The layout is shown in Fig.~\ref{fig:schema_v5}. A polarized neutron beam is divided at the beam splitter (BS) into two spatially separated paths, denoted by $I$ and $II$. Each path contains an active magnetic region before and after the central reflecting mirrors $C_I$ and $C_{II}$, and the two beams are subsequently recombined at the recombination element (REC) and detected at the output ports $H$ and $O$. A silicon phase shifter is placed along path $II$ and provides a controllable material-induced phase shift. In the sensitivity analysis considered here the magnetic fields $B_I$ and $B_{II}$ in the two arms are varied together, $B_I=B_{II}$, and the primary observable is the field-dependent reduction of the summed intensity $I_+ = I_H+I_O$, where $I_H$ and $I_O$ are the intensities at the two outputs.

	We describe the ordinary neutron $n$ and the hidden neutron $n'$ as a two-level system coupled by the mixing energy $\epsnn$. In the basis $(n,n')$ and for the spin branch with positive Zeeman shift, the effective Hamiltonian in a region with magnetic field $B$ and material potential $V$ is
	\begin{equation}
		M(B,V)=
		\begin{pmatrix}
			m_n+\mun B+V & \epsnn \\
			\epsnn & m_n+\dm
		\end{pmatrix},
		\label{eq:mass_hamiltonian}
	\end{equation}
	where $B$ is the magnetic field along the chosen quantization axis and $\mun=6.031\times10^{-8}\,\mathrm{eV/T}$ is the magnitude of the neutron magnetic moment. The quantity $V$ is the neutron optical potential, i.e., the coherent interaction energy of the neutron with a material; it is relevant for the silicon phase shifter discussed in Sec.~\ref{sec:phase_shifter}, while the active magnetic regions are modeled as vacuum propagation with $V=0$ (or $V\simeq0$ when a residual gas is present). The hidden state is assumed not to experience either the ordinary magnetic interaction or the material optical potential. The common rest energy $m_n$ contributes only an overall phase to the evolution. The energy detuning, i.e., the difference between the diagonal energies in the absence of mixing, is therefore
	$
	\Delta(B,V;\dm)=\mun B+V-\dm.
	$
	For a homogeneous region crossed in a time $t$, the $n\to n'$ conversion probability is
	\begin{equation}
		P_{n\to n'}(t;B,V)
		=
		\frac{4\epsnn^2}{\Delta^2+4\epsnn^2}
		\sin^2\!\left[
		\frac{t}{2\hbar}\sqrt{\Delta^2+4\epsnn^2}
		\right].
		\label{eq:transition_probability}
	\end{equation}
	In the active vacuum regions, where $V=0$, the diagonal energies of the ordinary and hidden states coincide at the resonance field
	$
	B_{\rm res}=\frac{\dm}{\mun}.
	$
	At resonance the mixing angle is maximal and Eq.~\eqref{eq:transition_probability} reduces to
	\begin{equation}
		P_{n\to n'}(t;\Delta=0)
		=
		\sin^2\!\left(\frac{t}{\tau_{nn'}}\right)
		\simeq
		\left(\frac{t}{\tau_{nn'}}\right)^2,
		\quad t\ll\tau_{nn'}.
		\label{eq:resonant_weak_limit}
	\end{equation}
	The numerical calculation uses two successive active propagation intervals of length $L=0.30\,\mathrm{m}$. The neutron component is reflected at the intermediate mirrors $C_I$ and $C_{II}$ and subsequently propagates through the second active interval, whereas the hidden component continues along its trajectory and is removed from the reflected beam in the model. Conversion therefore appears as a reduction of the ordinary-neutron intensity reaching the interferometer outputs. For a neutron with wave number $k=2\pi/\lambda$, where $\lambda$ is its wavelength, and output index $a=H,O$, let
	$I_a(B,k;\epsnn,\dm,\bm{\eta}_{\rm inst})$ denote the dimensionless detection probability at that output, normalized to the incident neutron intensity, where $\bm{\eta}_{\rm inst}$ denotes the relevant instrumental parameters. The total ordinary-neutron detection probability is $I_{\rm tot}(B,k)\equiv I_+(B,k)=I_H(B,k)+I_O(B,k)$, with the remaining arguments suppressed. These intensities describe the normalized optical response; the additional rate normalization and relative detection factors needed to predict measured counts are introduced in Sec.~\ref{sec:detectors}.

	The quantities $I_a$ depend on the neutron wavelength and, when magnetic-field variations are included, on the field sampled by the neutron. We denote by $\overline I_a$ the corresponding averaged detector intensity. The spectral and magnetic-field averages are introduced explicitly in Secs.~\ref{sec:broadband} and~\ref{sec:magnetic}. For a chosen mass splitting $\dm$, the resonant configuration is set to the resonance field defined above, $B_{\rm on}=B_{\rm res}$, while a detuned configuration
	$
	B_{\rm off}=B_{\rm on}+\Delta B_{\rm mod}
	$
	provides the reference measurement. The detuning $\Delta B_{\rm mod}$ is chosen from the calculated resonance profile so that conversion in the off configuration is strongly suppressed. The predicted conversion at the reference field is retained in the intensity ratio.

	The field-dependent disappearance observable is defined as
	$\Slock(\epsnn,\dm)=1-\overline I_{\rm tot}(B_{\rm on})/\overline I_{\rm tot}(B_{\rm off})$.
	A positive $\Slock$ corresponds to the additional loss of ordinary neutrons through resonantly enhanced $n\to n'$ conversion. The term magnetic lock-in refers here to the extraction of a signal correlated with the alternation of the two magnetic settings. A common multiplicative attenuation cancels in this ratio. A change of the instrumental response between the two fields, however, modifies the ratio in the same way as a disappearance signal and must be constrained by calibration.

	\section{Counting statistics}
	\label{sec:statistics}

	The disappearance observable introduced in Sec.~\ref{sec:model} is measured by comparing the neutron counts accumulated in the resonant and detuned magnetic configurations. Let $T_{\rm on}$ and $T_{\rm off}$ denote the corresponding counting times, and let
	$\Non=N_H^{\rm on}+N_O^{\rm on}$ and $\Noff=N_H^{\rm off}+N_O^{\rm off}$
	be the total counts summed over the two detector ports in the resonant and detuned reference configurations, respectively.

	The measured count rates are then
	$\widehat R_{\rm on}=\Non/T_{\rm on}$ and $\widehat R_{\rm off}=\Noff/T_{\rm off}$,
	and the corresponding estimator of the magnetic-lock-in observable is
	$\widehat{\Slock}=1-\widehat R_{\rm on}/\widehat R_{\rm off}=1-\Non T_{\rm off}/(\Noff T_{\rm on})$.

	Even for a stationary beam, the number of detected neutrons fluctuates between repeated acquisitions. Modeling these fluctuations by independent Poisson counts in the two magnetic configurations gives
	$\operatorname{Var}(\Non)=\mathbb{E}[\Non]$ and
	$\operatorname{Var}(\Noff)=\mathbb{E}[\Noff]$,
	where $\mathbb E$ and $\operatorname{Var}$ denote the expectation and variance at fixed exposure. Propagating the two independent counting errors through $\widehat{\Slock}$ and estimating the Poisson means by the observed counts gives the large-count statistical variance
	\begin{equation}
		\sigma^2_{\rm stat}(\Slock)
		=
		\left(\frac{T_{\rm off}}{T_{\rm on}}\right)^2
		\left[
		\frac{\Non}{\Noff^2}
		+
		\frac{\Non^2}{\Noff^3}
		\right].
		\label{eq:slock_stat_error}
	\end{equation}

	For the small disappearance signals considered here, the expected on- and off-resonance count rates are close to one another. Let $R_{\rm on}$ and $R_{\rm off}$ be the expected count rates summed over the two detectors, with common reference scale $R_{\rm ref}\simeq R_{\rm on}\simeq R_{\rm off}$. For equal accumulated counting times $T_{\rm on}=T_{\rm off}=T_{\rm int}$ and total exposure $T_{\rm tot}=2T_{\rm int}$, Eq.~\eqref{eq:slock_stat_error} reduces to
	\begin{equation}
		\sigma_{\rm stat}(\Slock)\simeq2/\sqrt{R_{\rm ref}T_{\rm tot}}.
	\end{equation}
	When counting statistics dominate over systematic uncertainties, a fractional deficit $\Slock$ has approximate Gaussian-equivalent statistical significance
	$
	Z_{\rm stat}\simeq \Slock\sqrt{R_{\rm ref}T_{\rm tot}}/2,
	$
	where $Z_{\rm stat}$ is expressed in units of standard deviations (for example, $Z_{\rm stat}=3$ corresponds to a $3\sigma$ statistical effect). The total exposure required for a target significance is therefore
	$
	T_{\rm tot}\simeq4Z_{\rm stat}^2/(R_{\rm ref}\Slock^2).
	$
	For a fixed total exposure, an equal division between the two magnetic configurations minimizes the leading statistical variance when $R_{\rm on}\simeq R_{\rm off}$. More generally, the optimal allocation satisfies approximately
	$T_{\rm on}/T_{\rm off}\simeq\sqrt{R_{\rm off}/R_{\rm on}}$.
	Since the disappearance signal is small in the region considered here, the equal-exposure approximation is sufficient. Alternating the resonant and detuned configurations on timescales short compared with instrumental drifts reduces their contribution to the measured ratio. The times $T_{\rm on}$ and $T_{\rm off}$ are the accumulated live times, i.e., the intervals during which counts are recorded in each configuration.

	For a locally tuned measurement we take the detected reference rate
$R_{\rm ref}=20\,\mathrm{s^{-1}}$, as motivated by the acceptance
calculation in Sec.~\ref{sec:high_rate}. The reference local exposure
is $T_{\rm tot}=30$ days of live time, equally divided between
$T_{\rm on}$ and $T_{\rm off}$.
	At $R_{\rm ref}=20\,\mathrm{s^{-1}}$, this corresponds to
	$
	N_{\rm ref}
	=
	R_{\rm ref}T_{\rm int}
	=
	2.59\times10^7
	$
	counts accumulated in each configuration and therefore to a statistics-only fractional precision
	$
	\sigma_{\rm stat}(\Slock)
	\simeq
	2.78\times10^{-4}.
	$

	In the weak-mixing regime, Eq.~\eqref{eq:resonant_weak_limit} implies that the disappearance signal scales as $\Slock\propto\tau_{nn'}^{-2}$.
	Since the statistics-only uncertainty decreases as $(R_{\rm ref}T_{\rm tot})^{-1/2}$, the accessible oscillation time $\tau_{nn'}^{\rm sens}$ at fixed statistical significance scales as
	$
	\tau_{nn'}^{\rm sens}
	\propto
	(R_{\rm ref}T_{\rm tot})^{1/4}.
	$
	Increasing the detected neutron rate or the exposure therefore improves the accessible oscillation time only through a fourth-root dependence. This scaling applies while counting statistics dominate; once relative on/off instrumental uncertainties become comparable to the statistical precision, additional exposure alone no longer produces the same improvement.

	\section{Broadband VCN spectrum and spectral averaging}
	\label{sec:broadband}

	The VCN beam is not monochromatic. Neutrons with different wave numbers
	have different velocities and therefore spend different times in the active
	magnetic regions. Since the $n\rightarrow n'$ conversion probability depends
	on this interaction time, the predicted intensity must be averaged over the
	neutron spectrum.

	We model the accepted wave-number distribution as
	\begin{align}
		f_k(k;k_0,\sigma_k)
		&\propto
		\exp\left[-\frac{(k-k_0)^2}{2\sigma_k^2}\right],
		\label{eq:gaussian_spectrum}\end{align}
	where $k_0=2\pi/\lambda_0$ is the central wave number and $\sigma_k$ is the standard deviation. We use
	$
	\lambda_0=40\,\text{\AA}
	$ and
	$
	\delta_k\equiv\frac{\sigma_k}{k_0}=0.10.
	$
	For a neutron of wave number $k$,
	$	v(k)=\frac{\hbar k}{m_n},
	$ and $
	t(k)=\frac{L}{v(k)}
	=\frac{Lm_n}{\hbar k},
	$
	where $v(k)$ is the neutron speed and $t(k)$ the interaction time for one active segment of length $L$. In the reduced calculation the spectrum is averaged over
	$
	0.2k_0\leq k\leq1.8k_0,
	$
	corresponding to $k_0\pm8\sigma_k$ for $\delta_k=0.10$. This interval contains essentially the full Gaussian weight. The averaged intensity is
	\begin{equation}
		\langle I_a(B)\rangle_k
		=
		\int_{0.2k_0}^{1.8k_0}
		\dd k\,
		f_k(k;k_0,\sigma_k)\,
		I_a(B,k),
		\quad a=H,O,
		\label{eq:spectral_average}
	\end{equation}
	where $f_k$ is normalized to unity over the integration interval and $\langle I_a(B)\rangle_k$ denotes the spectral average. The notation $\overline I_a$ includes the effective-field average as well when effective magnetic-field broadening is considered.

	Since $\lambda=2\pi/k$, a Gaussian distribution in $k$ is not exactly Gaussian in wavelength. For the relatively narrow spectrum considered here, however,
	$
	\frac{\sigma_\lambda}{\lambda_0}
	\simeq
	\frac{\sigma_k}{k_0}
	=
	\delta_k
	$
	to leading order, with $\sigma_\lambda$ the wavelength standard deviation. Both calculations use the same spectral model; a measured neutron spectrum can enter directly through the weight in Eq.~\eqref{eq:spectral_average}.

	\section{Magnetic-field calibration and resonance broadening}
	\label{sec:magnetic}

	The magnetic field determines both the position and the width of the resonant $n\to n'$ response. Because the active-vacuum resonance is set by the magnetic compensation condition introduced in Sec.~\ref{sec:model}, an uncertainty in the absolute magnetic-field scale translates into an uncertainty in the inferred mass splitting. We write the actual field $B_j^{\rm true}$ corresponding to a nominal setting $B_j$ as
	$
	B_j^{\rm true}=B_j+\delta B_0,
	$
	where the index $j$ labels a magnetic setting and $\delta B_0$ denotes a common calibration offset. A standard deviation $\sigma_{B_0}$ on this offset therefore corresponds to a mass-splitting uncertainty
	$
	\sigma_{\dm}^{(B)}=\mun\sigma_{B_0}.
	$
	Magnetometry supplies an independent constraint on $\delta B_0$, which enters the likelihood through a calibration term.

	Besides shifting the resonance center, field variations broaden the measured response because different neutrons encounter different detunings. In the reduced numerical model, the distribution of effective fields sampled by the detected neutrons at nominal setting $B_j$ is
	\begin{align}
		G_B(B';B_j+\delta B_0,\sigma_B^{\rm inh})
		={}&
		\frac{1}{\sqrt{2\pi}\,\sigma_B^{\rm inh}}
		\nonumber\\
		&\times
		\exp\!\left[
		-\frac{(B'-B_j-\delta B_0)^2}
		{2(\sigma_B^{\rm inh})^2}
		\right].
		\label{eq:field_kernel}
	\end{align}
	Here $\sigma_B^{\rm inh}$ is the root-mean-square (rms) width of the distribution and $B'$ is the effective field sampled by an individual neutron. We model quasi-static variations across trajectories or between events by assigning the same effective field to both active segments of a trajectory and a common distribution to both arms. We refer to the resulting ensemble-averaged resonance profile as effective magnetic-field broadening. This prescription describes the spread of quasi-static fields sampled by different neutrons. For a field varying along a given trajectory, the evolution through an active segment in arm $\alpha=I,II$ is instead governed by
	\begin{equation}
		\begin{aligned}
			U_\alpha(t)&=\mathcal T\exp\!\left[-\frac{i}{\hbar}
			\int_0^t M_\alpha(t')\,\dd t'\right],\\
			M_\alpha(t')&=M(B_\alpha(\bm x_\alpha(t')),V_\alpha(t')).
		\end{aligned}
		\label{eq:trajectory_evolution}
	\end{equation}
	where $\bm x_\alpha(t')$ is the neutron trajectory, $B_\alpha$ is the local field along the fixed spin quantization axis, $V_\alpha$ is the optical potential along that trajectory, and $\mathcal T$ denotes time ordering. Here $M$ is the selected-spin Hamiltonian of Eq.~\eqref{eq:mass_hamiltonian}, with the corresponding Zeeman sign for the other spin branch. The two segments are connected by the mirror operation described in Sec.~\ref{sec:model}. For fields whose direction varies, the evolution must also retain both spin states. This trajectory-dependent treatment determines both the conversion amplitudes and the relative ordinary-neutron phase at the recombiner, whereas the effective-field average isolates the broadening due to variations between trajectories or events.

	The intrinsic magnetic-field width of the resonance is set by the finite propagation time through an active region. In the weak-mixing limit, Eq.~\eqref{eq:transition_probability} can be written close to resonance as
	\begin{equation}
		P_{n\to n'}
		\simeq
		\left(\frac{\epsnn t}{\hbar}\right)^2
		\operatorname{sinc}^2\!\left[
		\frac{\mun(B-B_{\rm res})t}{2\hbar}
		\right],
		\label{eq:weak_field_profile}
	\end{equation}
	where $\operatorname{sinc}(x)=\sin x/x$. For one active segment of length $L=0.30\,\mathrm{m}$ and the central wavelength $\lambda_0=40\,\text{\AA}$, the corresponding neutron velocity is approximately
	$
	v_0\simeq99\,\mathrm{m\,s^{-1}},
	$
	giving an interaction time
	$
	t_L=\frac{L}{v_0}\simeq3.03\,\mathrm{ms}.
	$
	The corresponding magnetic detuning scale for this interaction time is
	$
	\Delta B_{\rm tof}
	\equiv
	\frac{\hbar}{\mun t_L}
	\simeq3.6\,\mu\mathrm{T}.
	$

	The half-maximum condition for the function $\operatorname{sinc}^2(x)$ occurs at $|x|\simeq1.39$. The full width at half maximum (FWHM) of the single-segment resonance is therefore
	$
	\Delta B_{\rm FWHM}
	\simeq
	5.57\,\Delta B_{\rm tof}
	\simeq20\,\mu\mathrm{T}.
	$
	Spectral averaging and the two-segment propagation determine the detector-level resonance profile from this intrinsic scale.

	For the reduced calculation we set the effective field width at each nominal setting $B_j$ through $\sigma_B^{\rm inh}=\beta_B|B_j|$, with relative widths $\beta_B=10^{-5}$ and $10^{-4}$.
	Over the mass interval $2\leq\dm/\mathrm{neV}\leq10$, corresponding to resonant fields of approximately $33$--$166\,\mathrm{mT}$, these assumptions imply absolute rms widths of approximately $0.3$--$1.7\,\mu\mathrm{T}$ and $3.3$--$16.6\,\mu\mathrm{T}$, respectively. The first width remains well below the intrinsic resonance width, whereas the second becomes comparable to it toward the upper end of the interval.

	\section{Interferometric channel and phase shifter}
	\label{sec:phase_shifter}

	The two output ports provide access to the coherence of the ordinary-neutron amplitudes in addition to their total intensity. We introduce the port difference $I_-\equiv I_H-I_O$. Let $a_I$ and $a_{II}$ denote the ordinary-neutron amplitudes reaching the recombination element REC from paths $I$ and $II$. For an ideal balanced and unitary recombiner,
	$I_H+I_O=|a_I|^2+|a_{II}|^2$,
	independently of the relative interferometric phase. By contrast, the port difference contains the interference term and can be written, up to the fixed phase convention of the recombiner, as
	$I_H-I_O=2\,\mathrm{Re}[e^{i\phi_{\rm REC}}a_I^*a_{II}]$,
	where $\phi_{\rm REC}$ denotes the relative phase associated with the reflection and transmission amplitudes of the recombination element. Thus the sum measures the surviving ordinary-neutron intensity, while the difference measures its interference.

	The silicon phase shifter located along path $II$ provides controlled access to this interferometric degree of freedom (see Ref.~\cite{Lemmel2010} for a discussion of neutron phase shifters). For an effective path length $D_{\rm eff}$ through the slab, the ordinary-neutron phase accumulated relative to propagation in the surrounding medium is
	$
	\chi(k_{\rm air})
	=
	k_{\rm air}D_{\rm eff}\,[n_r(k_{\rm air})-1],
	$
	where
	$
	n_r(k_{\rm air})
	=
	\sqrt{
		1-\frac{V_{\rm Si}-V_{\rm air}}
		{E-V_{\rm air}}
	}
	$
	is the neutron refractive index relative to the surrounding medium, with $E$ the neutron kinetic energy in vacuum, $V_{\rm Si}$ the silicon optical potential, and $V_{\rm air}$ the optical potential of the surrounding medium. Here $k_{\rm air}=\sqrt{2m_n(E-V_{\rm air})}/\hbar$ is the wave number in the surrounding medium, which becomes the vacuum wave number $k$ for $V_{\rm air}=0$. The silicon optical potential is
	$
	V_{\rm Si}
	=
	\frac{2\pi\hbar^2N_{\rm Si}b_c}{m_n},
	$
	where $N_{\rm Si}$ is the silicon atomic number density and $b_c$ is its coherent neutron scattering length. The reference geometry of Ref.~\cite{OriginalProposal} uses a silicon slab with physical thickness $D=200\,\mu\mathrm{m}$ and $V_{\rm Si}\simeq54\,\mathrm{neV}$; rotation of the slab changes $D_{\rm eff}$ and therefore allows the relative phase to be tuned.

	Varying $\chi$ changes the distribution of counts between the ports and allows the interference visibility (the contrast of the intensity modulation) and the operating phase to be determined. The phase shifter lies outside the active magnetic regions, leaving the resonance condition unchanged.

	Neutron--hidden-neutron mixing can modify both the amplitudes and the phases reaching the recombiner. To establish what the interferometric readout adds to the disappearance measurement, define the intrinsic port asymmetry $C_j=\overline I_{-,j}/\overline I_{+,j}$ at magnetic setting $j$, with $\overline I_{\pm,j}=\overline I_{H,j}\pm\overline I_{O,j}$ and $\overline I_{a,j}=\overline I_a(B_j)$. Suppose that the two ports have detection efficiencies $e_H=\bar e(1+d_{\rm e})$ and $e_O=\bar e(1-d_{\rm e})$, where $\bar e=(e_H+e_O)/2$ is their mean efficiency and $d_{\rm e}=(e_H-e_O)/(e_H+e_O)$ is their fractional imbalance. Let $J_{a,j}=e_a\overline I_{a,j}$ be the detected intensity at port $a$. The detected summed intensity $J_{+,j}=J_{H,j}+J_{O,j}$ is then
	\begin{equation}
		J_{+,j}=\bar e\,\overline I_{+,j}(1+d_{\rm e} C_j).
		\label{eq:port_imbalance}
	\end{equation}
	Even in the absence of disappearance, a change in the interferometric working point between the two magnetic settings can thus produce an apparent deficit if the port efficiencies differ. For a stable efficiency imbalance $d_{\rm e}$, the fractional deficit arising solely from this redistribution is
	\begin{equation}
		S_{\rm port}=-\frac{d_{\rm e}(C_{\rm on}-C_{\rm off})}{1+d_{\rm e} C_{\rm off}}.
		\label{eq:port_spurious}
	\end{equation}
	In the limit $|d_{\rm e}|\ll1$, the requirement that this effect remain below a target fractional uncertainty $\sigma_{\rm rel}$ on the relative on/off response becomes $|d_{\rm e}(C_{\rm on}-C_{\rm off})|\lesssim\sigma_{\rm rel}$.

	For negligible background, the expected measured port asymmetry at fixed nonzero total count is $C_j^{\rm det}=(J_{H,j}-J_{O,j})/J_{+,j}=(C_j+d_{\rm e})/(1+d_{\rm e}C_j)$. The intrinsic asymmetry can therefore be recovered from the measured one using the calibrated efficiency imbalance. For balanced detection, the conditional distribution of the count $N_{H,j}$ at fixed total count $N_j=N_{H,j}+N_{O,j}$ is binomial with probability $(1+C_j)/2$ for an individual count to appear at port $H$. The asymmetry estimator $(N_{H,j}-N_{O,j})/N_j$ consequently has conditional variance $(1-C_j^2)/N_j$. The port asymmetry thus measures changes in the interference pattern without requiring the absolute intensity normalization. With independently calibrated port efficiencies, this information can be used to diagnose or correct the contribution in Eq.~\eqref{eq:port_spurious}. Section~\ref{sec:port_budget} quantifies the resulting correction
and its contribution to the sensitivity.

	In the lossless limit, collecting both outputs retains the incident intensity in the absence of conversion. The sensitivity projections use the disappearance channel at the specified detected rate; the phase-sensitive readout constrains redistribution between the outputs through the joint likelihood developed in Sec.~\ref{sec:covariance}.

	Field-independent attenuation by the phase shifter enters the common rate normalization, while modulation-correlated variations of its transmission enter the relative response.

	\section{Polarization and experimental normalization}
	\label{sec:detectors}

	The magnetic resonance depends on the neutron spin because the two spin projections experience opposite Zeeman shifts. We denote by $P$ the fraction of the beam in the spin state for which the ordinary-neutron shift is $+\mun B$. The opposite fraction $1-P$ experiences the detuning with the opposite magnetic sign. A spin-averaged detector response can therefore be written as
	$I_a^{\rm pol}=P\,I_a^{(+)}+(1-P)\,I_a^{(-)}$, with $a=H,O$,
	where $I_a^{(\pm)}$ denotes the response of port $a$ for the two spin branches, whose active-region detunings are
	$
	\Delta_\pm=\pm\mun B-\dm.
	$
	The spin populations can be characterized by neutron polarimetry~\cite{MuPAD}. In the reduced calculation we use a selected-spin fraction $P_0=0.99$. The conventional beam polarization is the population asymmetry $\mathcal P=(N_+-N_-)/(N_++N_-)=2P_0-1=0.98$, where $N_+$ and $N_-$ are the populations of the two spin branches. The complete interferometer calculation uses the selected spin branch, whereas the reduced calculation explicitly retains the finite $1-P_0$ population through the spin average defined above. To relate the disappearance signal to the measured counts, we account for the neutron flux and detector response in each magnetic configuration. The predicted signal rate at port $a$ and setting $j$ is
	$r_{aj}^{\rm sig}=\mathcal N\,F_j\,\kappa_{aj}\,\overline I_a(B_j)$,
	where $\mathcal N$ is a common rate normalization (in counts per unit time), $F_j$ is the incident flux relative to its reference value, and $\kappa_{aj}$ describes the relative transport, collection, and detection efficiencies. Both $F_j$ and $\kappa_{aj}$ are dimensionless factors, defined relative to nominal conditions. Here $\overline I_a$ includes the spin average appropriate to the calculation. The normalization $\mathcal N$ sets the absolute rate scale, while the factors $F_j$ and $\kappa_{aj}$ describe deviations from the reference conditions.

	An independent neutron monitor, which measures the incident beam intensity, can constrain the flux factor $F_j$. Let $R_{\rm mon}$ be its expected signal rate at the reference flux. For monitor live time equal to the counting time $T_j$ and accepted monitor background rate $b_{\rm mon}$, the expected monitor counts are
	$\mu_j^{\rm mon}=T_j(F_jR_{\rm mon}+b_{\rm mon})$.
	The measured monitor count is denoted by $M_j$.

	The monitor constrains relative flux changes between magnetic settings, which could otherwise imitate conversion.
	For detector port $a\in\{H,O\}$ and magnetic setting $j$, the total expected count rate is $r_{aj}^{\rm obs}=r_{aj}^{\rm sig}+b_{aj}$. Here $r_{aj}^{\rm sig}$ includes the loss due to $n\to n'$ conversion, while $b_{aj}$ is the rate of background events accepted by the same event-selection criteria.
	A constant accepted background dilutes the disappearance contrast. To quantify this effect, consider equal on/off live times, a reference signal rate $R$, and a stationary Poisson background rate $b=\sum_{a=H,O}b_{aj}$, independent of the setting $j$ and known from auxiliary measurements. A fractional signal deficit $S$ changes the signal rate from $R$ to $R(1-S)$, so the fractional deficit of the total rate becomes $S_{\rm obs}=S/(1+b/R)$. The variance of the background-subtracted fractional signal deficit in the small-signal limit is then
	\begin{equation}
		\sigma^2_{\rm stat,b}\simeq\frac{4(1+b/R)}{RT_{\rm tot}}.
		\label{eq:background_variance}
	\end{equation}
	Thus, since the weak-mixing deficit scales as $\tau_{nn'}^{-2}$, the statistics-limited sensitivity obeys $\tau_{95}(b)/\tau_{95}(0)=(1+b/R)^{-1/4}$ at fixed signal rate and exposure. An uncertain difference of the accepted background rates between the two settings, with standard deviation $\sigma_{\Delta b}$, adds a fractional uncertainty $\sigma_{\Delta b}/R$ to the relative response budget. For independent calibration and background uncertainties their variances add. Time-dependent backgrounds or departures from Poisson counting require their own covariance model. The numerical projections below take negligible accepted background and detector dead time (the interval during which a detector cannot register another event), with the exposure expressed in live time.

	\section{Statistical inference and projected sensitivity}
	\label{sec:covariance}

	To translate the predicted conversion probability into a limit on the mixing energy, we evaluate the likelihood of the measured counts for each oscillation hypothesis. The Poisson likelihood retains the counting information underlying the disappearance ratio. For detector port $a\in\{H,O\}$ and magnetic setting $j$, the expected number of counts is obtained by multiplying the accumulated counting time $T_j$ by the predicted mean count rate $r_{aj}$:
	\mbox{$\mu_{aj}(\epsnn,\dm;\bm{\eta})=T_j\,r_{aj}(\epsnn,\dm;\bm{\eta})$},
	where $r_{aj}=r_{aj}^{\rm obs}$ is the expected count rate, including any accepted background, and $\bm{\eta}$ collects the instrumental parameters of the analysis.
	Let $N_{aj}$ denote the measured count in the same port and setting. At fixed predicted mean $\mu_{aj}$, its Poisson probability is $\mu_{aj}^{N_{aj}}e^{-\mu_{aj}}/N_{aj}!$. The likelihood is the product of these probabilities, regarded as a function of the model parameters at fixed observed counts. For independent counting channels, the deviance (minus twice the logarithm of the likelihood relative to that of a model reproducing every observed count) is
	\begin{equation}
		\chi^2_{\rm sig}
		=
		2\sum_{a,j}
		\left[
		\mu_{aj}-N_{aj}
		+
		N_{aj}\ln\!\left(\frac{N_{aj}}{\mu_{aj}}\right)
		\right],
		\label{eq:poisson_deviance}
	\end{equation}
	with the logarithmic term defined to vanish for $N_{aj}=0$. Auxiliary monitor counts can be included through an analogous Poisson term, while independently calibrated instrumental quantities can be allowed to vary within their measured calibration uncertainties. For the numerical projections we use the signal counts to infer the common normalization and represent the auxiliary calibration information by constraints on the instrumental parameters, rather than simulating monitor counts separately.

	The two-port likelihood can be separated into a disappearance term and an interferometric term. At magnetic setting $j$, define $N_j=N_{H,j}+N_{O,j}$, $\mu_j=\mu_{H,j}+\mu_{O,j}$ and $p_j=\mu_{H,j}/\mu_j$. The product of the two independent Poisson probabilities is exactly a Poisson probability for $N_j$ with mean $\mu_j$, times a binomial probability for $N_{H,j}$ conditional on $N_j$, with success probability $p_j$. Accordingly, Eq.~\eqref{eq:poisson_deviance} separates into the summed-count deviance and the conditional port deviance
	\begin{equation}
		\chi^2_{\rm port}=2\sum_j\left[N_{H,j}\ln\!\tfrac{N_{H,j}}{N_jp_j}+N_{O,j}\ln\!\tfrac{N_{O,j}}{N_j(1-p_j)}\right].
		\label{eq:port_deviance}
	\end{equation}
	The same zero-count convention as in Eq.~\eqref{eq:poisson_deviance} applies, and a setting with $N_j=0$ contributes no conditional information. For negligible background and the relative port response of Sec.~\ref{sec:phase_shifter}, $p_j=(1+C_j^{\rm det})/2$. The summed-count term measures disappearance, whereas the conditional term measures the redistribution of intensity independently of the common normalization. The estimator $\widehat{\Slock}$ is a representation of the summed counts, not an additional measurement. The numerical disappearance projections use the summed-count term and the adopted residual calibration constraints.

	To separate mixing from instrumental variations, we minimize the deviance over the instrumental parameters for each tested oscillation hypothesis. This procedure, known as profiling, gives
	\begin{equation}
		\chi^2_{\rm prof}(\epsnn,\dm)=\min_{\bm\eta}\left[
		\chi^2_{\rm sig}(\epsnn,\dm;\bm\eta)+\chi^2_{\rm cal}(\bm\eta)\right],
	\end{equation}
	where $\chi^2_{\rm cal}$ encodes the likelihood of the auxiliary calibration measurements. A shift of an instrumental parameter can compensate part of the predicted disappearance, but such a shift is penalized according to its calibration uncertainty. Profiling therefore tests each mixing hypothesis at the instrumental response most compatible with both sets of measurements.

	To forecast the sensitivity in the absence of mixing we use an Asimov data set~\cite{Cowan2011}: the counts are set equal to their expectation under the no-mixing hypothesis,
	$N_{aj,A}=\mu_{aj}(\epsnn=0,\dm;\bm\eta_0)$,
	where $\bm\eta_0$ denotes the nominal calibrated instrumental parameters. The auxiliary data are likewise set to their nominal expectations. This construction removes statistical fluctuations from the forecast. For these data the profiled test statistic is
	\begin{equation}
		Q_A(\epsnn,\dm)=\chi^2_{\rm prof}(\epsnn,\dm)-\chi^2_{\rm prof,min},
	\end{equation}
	where $\chi^2_{\rm prof,min}$ is obtained by minimizing also over the mixing energy at fixed $\dm$. The statistic quantifies the disagreement between the tested mixing hypothesis and the expected no-mixing data, after accounting for the allowed instrumental variations. We use $q_A$ for the count-level Asimov deviance with calibration parameters held at their nominal values and $Q_A$ after profiling the calibration shifts. The common count normalization is profiled in both cases.

	The effect of the conditional port information can be illustrated by linearizing the on/off response around a balanced working point, with a small, independently calibrated and stable imbalance $d_{\rm e}$. Write $x=C_{\rm on}-C_{\rm off}$ for the change of intrinsic asymmetry, and let $\sigma_x$ denote its statistical uncertainty obtained from the conditional port counts. With $S$ the physical fractional disappearance, the summed-rate deficit is $S-d_{\rm e}x-\eta$, where $\eta$ denotes the remaining relative signal-response shift, constrained by an independent Gaussian uncertainty $\sigma_\eta$. In the large-count, weak-signal limit, the local Asimov data with nominal $x=\eta=0$ give the joint quadratic deviance
	\begin{equation}
		\begin{aligned}[b]
			Q_{\rm joint}(S)
			&\simeq\min_{x,\eta}\left[
			  \frac{(S-d_{\rm e}x-\eta)^2}{\sigma_{\rm stat}^2}
			  +\frac{x^2}{\sigma_x^2}+\frac{\eta^2}{\sigma_\eta^2}
			\right]\\
			&=\frac{S^2}{\sigma_{\rm stat}^2+d_{\rm e}^2\sigma_x^2+\sigma_\eta^2}.
		\end{aligned}
		\label{eq:joint_port_profile}
	\end{equation}
	Here $\sigma_{\rm stat}$ is the summed-rate statistical uncertainty of Sec.~\ref{sec:statistics}. The factorization of the Poisson likelihood implies a vanishing mixed Fisher-information term between the total mean and the port probability (the expected curvature of the negative log likelihood). The quadratic approximation can therefore use the total and the conditional port fraction as separate statistical constraints. If $x$ were unconstrained, a nonzero $d_{\rm e}$ would allow intensity redistribution to mimic the disappearance in this linearized model. The conditional port measurement replaces this degeneracy with the finite variance $d_{\rm e}^2\sigma_x^2$. A common change in incident flux or in the transmission of both ports, by contrast, remains degenerate with disappearance and must be constrained by independent calibration. Equation~\eqref{eq:joint_port_profile} describes the calibration information obtained from the asymmetry without assigning it an additional mixing signal. The residual $\sigma_{\rm rel}$ used below describes the overall relative response after such corrections, and we do not add the illustrative port variance a second time.

	At fixed $\dm$, we increase the mixing energy from zero until $Q_A=2.71$, the conventional asymptotic one-sided 95\% threshold for a single nonnegative signal strength (proportional to $\epsnn^2$ in the weak-mixing regime). Denoting this first positive solution by $\epsnn^{95}$, we define the projected oscillation-time sensitivity as $\tau_{95}=\hbar/\epsnn^{95}$.
	Since $\tau_{nn'}$ is inversely proportional to the mixing energy, a larger $\tau_{95}$ corresponds to sensitivity to weaker mixing. We quote the asymptotic median expected exclusion sensitivity at each tested splitting, on the weak-mixing branch. This is a pointwise limit rather than the significance of discovering a resonance somewhere in a scan.

	\section{Projected sensitivity and systematic requirements}
	\label{sec:updated_sensitivity}

	We use the positive-splitting branch in the reference mass interval
	$
	2\leq\dm/\mathrm{neV}\leq10,
	$
	corresponding to resonant fields of approximately $33$--$166\,\mathrm{mT}$. For the local sensitivity considered in this section, the full reference exposure is assigned to the on/off pair at each tested splitting,
	$
	T_{\rm tot}=30~\mathrm{days},
	$ and $
	T_{\rm on}=T_{\rm off}=15~\mathrm{days}.
	$
	In the reduced vacuum model, the opposite-sign branch is obtained by reversing the selected spin relative to the field. The corresponding local projections are therefore represented on the $|\dm|$ axis in Fig.~\ref{fig:updated_vcn_reach}(a).

	\subsection{Full-interferometer calculation}

	We first evaluate the sensitivity with fixed instrumental calibration using the complete, summed interferometer signal defined in Sec.~\ref{sec:model}. At fixed wavelength, the calculation includes the beam splitter, central mirrors, recombiner, guide regions, and silicon phase-shifter propagation. The wavelength dependence is averaged over the broadband VCN spectrum described in Sec.~\ref{sec:broadband}.

	For each tested splitting we choose
	$B_{\rm on}=B_{\rm res}$ and $B_{\rm off}=B_{\rm on}+100\,\mu\mathrm{T}$.
	The intensity ratio includes the calculated conversion at both fields.

	With fixed relative calibration, the incident flux need not be known absolutely: the two Poisson count samples share a single unknown rate normalization. Let $I_{0,\rm on}$ and $I_{0,\rm off}$ be the intensities obtained from the complete interferometer calculation in the null hypothesis, and let $I_{\rm on}$ and $I_{\rm off}$ denote the corresponding summed, spectrally averaged intensities at the tested mixing. With $R$ denoting the expected detected rate in the off-resonance configuration under the no-mixing hypothesis, and with equal counting time $T_{\rm int}$,
	$N_{\rm off,0}=RT_{\rm int}$ and $N_{\rm on,0}=RT_{\rm int}I_{0,\rm on}/I_{0,\rm off}$.
	The corresponding fractions are defined by
	$p_0=N_{\rm on,0}/(N_{\rm on,0}+N_{\rm off,0})$ and $p=I_{\rm on}/(I_{\rm on}+I_{\rm off})$.
	Thus $p_0$ and $p$ describe the fraction of the total expected counts in the resonant configuration without and with mixing, respectively.

	To remove the dependence on the absolute flux normalization, for the tested-mixing hypothesis we write the expected counts as
	$
	\mu_{\rm on}=\nu p,$
	and $
	\mu_{\rm off}=\nu(1-p),
	$
	where $\nu$ is the total expected count in the two configurations and is allowed to vary freely. For the Asimov data generated under the no-mixing hypothesis,
	$
	n_{\rm on}=N_{\rm tot}p_0,
	$ and $
	n_{\rm off}=N_{\rm tot}(1-p_0),
	$
	with \(N_{\rm tot}=N_{\rm on,0}+N_{\rm off,0}\). Profiling the joint Poisson likelihood over \(\nu\) gives
	$
	\widehat{\nu}=n_{\rm on}+n_{\rm off}=N_{\rm tot}.
	$
	The overall number of counts fixes the normalization, leaving the relative population of the two samples to constrain the mixing. The quantities $p_0$ and $p$ therefore contain the relevant signal information. Substitution of the profiled normalization gives the two-configuration Asimov deviance
	\begin{equation}
		q_A
		=
		2N_{\rm tot}
		\left[
		p_0\ln\left(\frac{p_0}{p}\right)
		+
		(1-p_0)
		\ln\left(\frac{1-p_0}{1-p}\right)
		\right],
		\label{eq:fullmodel_shape_likelihood}
	\end{equation}
	The fixed-calibration full-interferometer reach follows by setting this statistic equal to the threshold introduced in Sec.~\ref{sec:covariance}.

	In the weak-mixing regime the expected scaling is $q_A\propto\epsnn^4$. The numerical calculation verifies this behavior at representative masses before using it to rescale the sensitivity with rate and exposure. For $R=20\,\mathrm{s^{-1}}$ the resulting sensitivity is of order
	$
	\tau_{95}\simeq0.20\,\mathrm{s}
	$
	throughout the explicitly calculated interval. The same calculation determines the detected-rate requirement for a fixed-calibration target $\tau_{95}=1\,\mathrm{s}$ from the least favorable point in the interval, giving approximately
	$
	R_{\rm det}^{\rm req}
	\simeq1.24\times10^4\,\mathrm{s^{-1}}.
	$

	\subsection{Reduced calculation of magnetic and calibration effects}

	The effective magnetic-field broadening and relative-calibration requirements are governed primarily by the survival probability in the active regions. We isolate these effects with a reduced model that retains the two successive conversion segments and the spin and spectral distributions. Let $p_\pm(B,k)$ be the single-segment transition probability of Eq.~\eqref{eq:transition_probability}, using the flight time of Sec.~\ref{sec:broadband}, $V=0$, and the spin-dependent detunings defined in Sec.~\ref{sec:detectors}. Since the central mirrors remove the hidden component produced in the first segment, two equal active segments give the ordinary-neutron survival factor $[1-p_\pm(B,k)]^2$. Including the finite selected-spin fraction, the averaged ordinary-neutron survival response in magnetic configuration $j$, with $j=\mathrm{on},\mathrm{off}$, is given by
	\begin{equation}
		\begin{aligned}
			g_j(\tau_{nn'},\dm,\delta B_0)
			=
			\Big\langle
			&P_0[1-p_+(B,k)]^2\\
			&+(1-P_0)[1-p_-(B,k)]^2
			\Big\rangle_j,
		\end{aligned}
		\label{eq:active_response}
	\end{equation}
	where the brackets denote the average over $k$ and, when effective magnetic-field broadening is included, over the effective field at nominal setting $B_j$, using the normalized weights of Eqs.~\eqref{eq:spectral_average} and~\eqref{eq:field_kernel}.
	In this calculation,
	$
	B_{\rm off}
	=
	B_{\rm on}
	+
	\max\left(
	100\,\mu\mathrm{T},
	8\beta_B B_{\rm on}
	\right)
	$.
	The field displacement increases with the broadening width $\beta_B B_{\rm on}$ to keep the reference outside the main resonance peak. The corresponding reduced-model deficit is
	$S=1-g_{\rm on}/g_{\rm off}$.

	Since $g_j=1$ in the absence of mixing, the two configurations have equal expected null counts $N=RT_{\rm int}$. Profiling their common normalization as in the full calculation gives the exact two-count Asimov deviance
	\begin{equation}
		q_A(S)
		=
		2N
		\ln\left[
		\frac{(2-S)^2}{4(1-S)}
		\right].
		\label{eq:updated_poisson_ratio}
	\end{equation}
	For $S\ll1$, Eq.~\eqref{eq:updated_poisson_ratio} reduces to $q_A\simeq R T_{\rm tot}S^2/4=Z_{\rm stat}^2$, consistently reproducing the counting-statistics estimate of Sec.~\ref{sec:statistics}.
	In the weak-mixing regime,
	$S=A_2(\dm)/\tau_{nn'}^2+\mathcal O(\tau_{nn'}^{-4})$, where $A_2(\dm)=\lim_{\tau_{nn'}\to\infty}\tau_{nn'}^2S$ is the coefficient that sets the leading disappearance strength in the weak-mixing limit; since $S$ is dimensionless, $A_2$ has dimensions of time squared. For the representative point at $2\,\mathrm{neV}$,
	$
	A_2(2\,\mathrm{neV})
	=
	1.8733\times10^{-5}\,\mathrm{s^2}.
	$
	For $R=20\,\mathrm{s^{-1}}$ and $15$ days per configuration, the reduced calculation gives $\tau_{95}=0.2024\,\mathrm{s}$ at $2\,\mathrm{neV}$, consistent with the scale of the complete-interferometer result.
	To separate the loss of sensitivity due to the broader resonance from that due to uncertainty in the measured rates, we compare two fixed-calibration cases, $\beta_B=10^{-5}$ and $10^{-4}$. In the latter case we then allow a residual relative on/off response uncertainty $\sigma_{\rm rel}=10^{-3}$ and a common field-offset uncertainty $\sigma_{B_0}=2\,\mu\mathrm{T}$.

	The resulting $30$-day sensitivities at the ILL/PF2 reference rate are summarized in Table~\ref{tab:reduced_systematics}.

	\begin{table}[t]
		\caption{Reduced-model local 95\% sensitivities $\tau_{95}$, in seconds, for $R=20\,\mathrm{s^{-1}}$ and $30$ days total exposure. ``Fixed calibration'' uses $\beta_B=10^{-5}$ with fixed calibration; ``Field broadened'' uses $\beta_B=10^{-4}$ with fixed calibration; ``Calibration included'' additionally allows the relative response and common field offset to vary within $\sigma_{\rm rel}=10^{-3}$ and $\sigma_{B_0}=2\,\mu\mathrm{T}$.\label{tab:reduced_systematics}}
		\begin{ruledtabular}
			\begin{tabular}{cccc}
				$\dm$ [neV] &
				\shortstack{Fixed\\calibration} &
				\shortstack{Field\\broadened} &
				\shortstack{Calibration\\included}\\
				\hline
				2  & 0.2024 & 0.1953 & 0.1011\\
				5  & 0.2020 & 0.1698 & 0.0879\\
				10 & 0.2006 & 0.1345 & 0.0696\\
			\end{tabular}
		\end{ruledtabular}
	\end{table}

	The reduced model also provides an independent check of the high-rate requirement. The detected rates required for $\tau_{95}=1\,\mathrm{s}$ at $\dm=2$, $5$, and $10\,\mathrm{neV}$ are approximately $1.19\times10^4$, $1.20\times10^4$, and $1.24\times10^4\,\mathrm{s^{-1}}$, respectively, in the fixed-calibration scenario. The weak mass dependence follows from magnetic tuning: at resonance, the conversion probability depends on the propagation time and mixing strength but not explicitly on $\dm$. The residual dependence comes from the finite effective-field width and the detuned response. The least favorable point reproduces the $\mathcal O(1.24\times10^4\,\mathrm{s^{-1}})$ design scale obtained from the complete-interferometer calculation.

	\subsection{Calibration requirement}

	We describe a residual mismatch between the mean on- and off-resonance
	responses by a multiplicative parameter $\eta_{\rm rel}$, with uncertainty
	$\sigma_{\rm rel}$. The uncertainty $\sigma_{\rm rel}$ specifies the precision of the relative rate calibration after flux, transport, and detector corrections. For a disappearance
	signal $S$, the corresponding effective deficit is
	$
	S_{\rm eff}
	=
	1-(1+\eta_{\rm rel})(1-S).
	$
	Thus, $\eta_{\rm rel}=0$ corresponds to perfectly matched on/off
	responses, while, for example, $\sigma_{\rm rel}=10^{-3}$ represents a
	residual relative-calibration uncertainty of $0.1\%$.

	The competition between counting statistics and relative calibration is particularly transparent in the weak-mixing limit at fixed absolute field calibration. To leading order, $S_{\rm eff}\simeq S-\eta_{\rm rel}$, so the Poisson deviance is approximately $(S-\eta_{\rm rel})^2/\sigma_{\rm stat}^2$, with $\sigma_{\rm stat}^2=4/(RT_{\rm tot})$. Including the independent Gaussian calibration term $\eta_{\rm rel}^2/\sigma_{\rm rel}^2$ and minimizing their sum over $\eta_{\rm rel}$ gives
	$
	Q_A
	\simeq
	\frac{\left(A_2/\tau_{nn'}^2\right)^2}
	{4/(RT_{\rm tot})+\sigma_{\rm rel}^2},
	$
	which gives
	$
	\tau_{95}
	\simeq
	\sqrt{A_2}
	\left\{
	2.71
	\left[
	\frac{4}{RT_{\rm tot}}+\sigma_{\rm rel}^2
	\right]
	\right\}^{-1/4}.
	$
	Here $4/(RT_{\rm tot})$ is the contribution from counting statistics,
	whereas $\sigma_{\rm rel}^2$ describes the residual uncertainty in the
	relative on/off response. This expression therefore
	makes explicit how the sensitivity scales with event rate, exposure, and
	relative-calibration precision. For a known stationary background, the counting term is replaced by Eq.~\eqref{eq:background_variance}, with any independently measured uncertainty in the background difference included in the residual calibration budget.

	In the infinite-statistics limit, the counting term vanishes and a target
	oscillation time $\tau_*$ can be reached only if
	$
	\sigma_{\rm rel}
	<
	\frac{A_2}{\sqrt{2.71}\,\tau_*^2}.
	$
	Across the calculated interval, the reduced response with $\beta_B=10^{-5}$ implies a maximum relative-calibration uncertainty of approximately $(1.10$--$1.14)\times10^{-5}$ for $\tau_*=1\,\mathrm{s}$. This is the limiting calibration requirement even with arbitrarily many detected neutrons.

	At finite exposure, the counting variance reduces the allowed calibration uncertainty for a given target sensitivity. Section~\ref{sec:port_budget} evaluates this budget for
	the locally tuned high-rate example.

	In the numerical calculation the absolute field offset $\delta B_0$ also changes the predicted conversion through the effective-field average. The deficit including both calibration shifts is
	\begin{equation}
		\begin{aligned}
			S_{\rm eff}(\tau_{nn'},\dm;\eta_{\rm rel},\delta B_0)
			&=
			1-(1+\eta_{\rm rel})\times\\
			&\left[
			1-S(\tau_{nn'},\dm;\delta B_0)
			\right].
			\label{eq:Seff_profiled}
		\end{aligned}
	\end{equation}
	For each tested oscillation hypothesis $(\tau_{nn'},\dm)$, we profile
	simultaneously over the relative-response mismatch and the magnetic-field
	offset,
	\begin{equation}
		\begin{aligned}
			Q_A(\tau_{nn'},\dm)
			=
			\min_{\eta_{\rm rel}>-1,\,\delta B_0}
			\Bigg[
			&q_A\!\left(
			S_{\rm eff}(\tau_{nn'},\dm;
			\eta_{\rm rel},\delta B_0)
			\right)
			\\
			&+
			\frac{\eta_{\rm rel}^2}{\sigma_{\rm rel}^2}
			+
			\frac{\delta B_0^2}{\sigma_{B_0}^2}
			\Bigg].
		\end{aligned}
		\label{eq:updated_relative_profile}
	\end{equation}

	The Gaussian terms represent independent auxiliary constraints on the two calibration shifts: a displacement by one standard deviation contributes one unit to the deviance. The condition $\eta_{\rm rel}>-1$ ensures a positive response factor. The minimum determines $\tau_{95}$ through the same criterion as in the fixed-calibration calculation.

	We verified the stability of the spectral and magnetic-field averages
	against increased numerical-integration accuracy. We also compared the
	exact finite-mixing sensitivity with its weak-mixing approximation, finding relative agreement better than $5\times10^{-4}$ for the reference-interval scenarios tested. For the local calculation, the weak-mixing
	expression exposes the sensitivity scaling,
	whereas the numerical limits in Table~\ref{tab:reduced_systematics} are obtained from the exact
	finite-mixing calculation with the instrumental parameters profiled according
	to Eq.~\eqref{eq:updated_relative_profile}.

	The dependence of the local disappearance sensitivity on the mass
	splitting can be understood from the resonant and reference responses.
	At small splittings, the resonance approaches zero magnetic field,
	while the reference field remains displaced by at least
	$100\,\mu\mathrm{T}$. The corresponding Zeeman-energy difference,
	approximately $6\,\mathrm{peV}$, exceeds the intrinsic energy scale
	$\hbar/t_L$ of a single active segment. The contrast between the two
	configurations is thus preserved in the reduced vacuum model, including
	both spin populations. The optical potential of a material can likewise
	detune the ordinary and hidden neutron states, as relevant to
	regeneration searches in matter~\cite{MM6,StasserMURMUR2021,AlmazanSTEREO2022}.
	Maintaining this sensitivity at low fields requires control of residual
	magnetic fields and optical potentials in the active regions.

	At larger splittings, the resonant field increases, and for a fixed
	relative field width $\beta_B$ the absolute field spread increases as
	well. The resulting reduction of the conversion response follows from
	Eq.~\eqref{eq:weak_field_profile}. For a single active segment with
	interaction time $t$, averaging over the Gaussian field distribution
	centered at resonance yields
	\begin{align}
		\langle p_+\rangle_B &\simeq
		\frac{t^2}{\tau_{nn'}^2}F(q),\nonumber\\
		F(q)&=\sqrt{\frac{\pi}{2}}\frac{\operatorname{erf}(\sqrt{2}q)}{q}
		-\frac{1-e^{-2q^2}}{2q^2},\nonumber\\
		q&=\frac{\beta_B|\dm|t}{2\hbar},
		\label{eq:broad_extension}
	\end{align}
	where $q$ is the dimensionless field-spread parameter, $F(q)$ is the
	reduction of the resonant conversion probability, $\operatorname{erf}$
	denotes the error function, and $F(0)=1$. For $q\gg1$,
	$F(q)\simeq\sqrt{\pi/2}/q$. Hence, at fixed detected rate, exposure,
	and relative calibration uncertainty, the local sensitivity approaches
	$\tau_{95}\propto|\dm|^{-1/2}$ when the reference remains sufficiently
	detuned. This asymptotic behavior describes the effect of effective
	magnetic-field broadening, whereas the maximum operating field sets
	the endpoint of the resonant search.

\begin{figure*}[t]
 \centering
 \IfFileExists{vcn_scan_reach.pdf}{%
  \includegraphics[width=\textwidth]{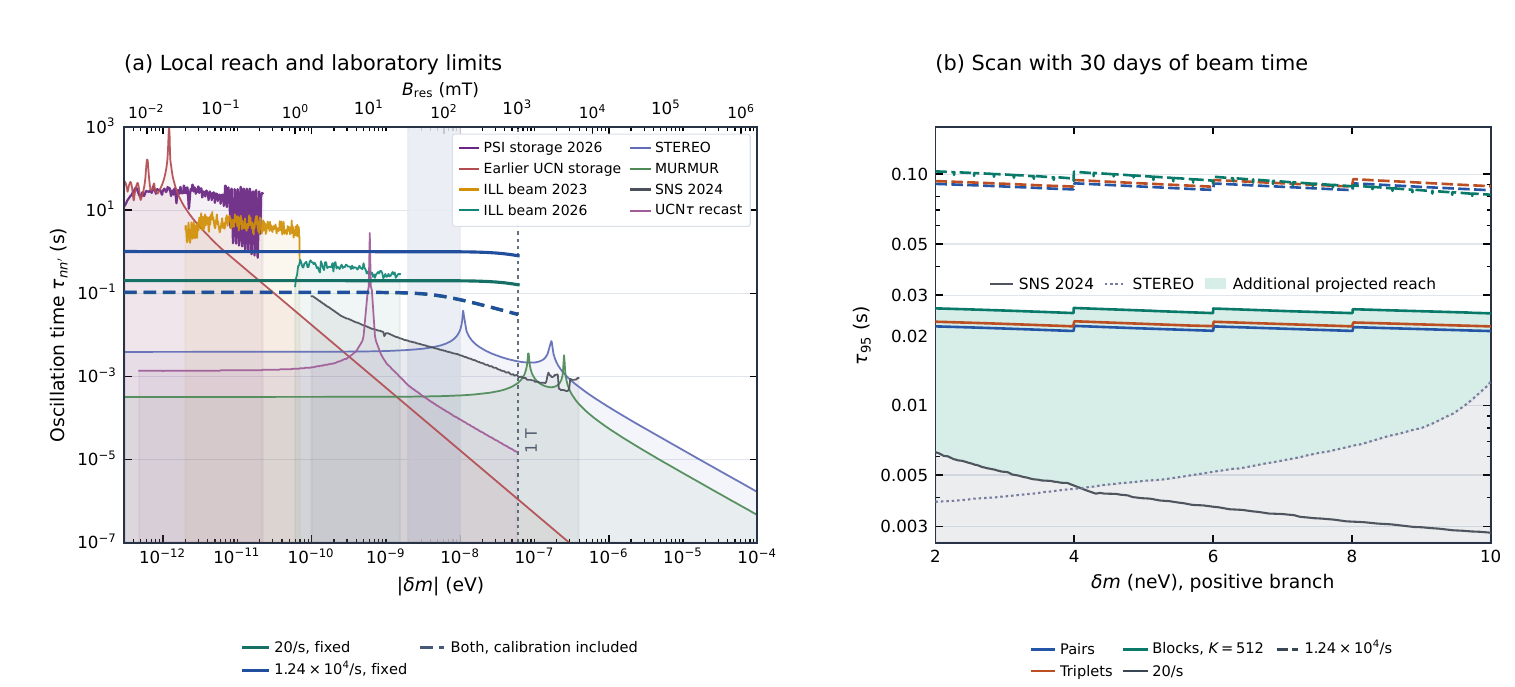}%
 }{%
  \includegraphics[width=\textwidth]{vcn_scan_reach(1).pdf}%
 }
\caption{Projected sensitivity to neutron--hidden-neutron
	oscillations compared with existing laboratory exclusion
	regions.
	(a) Local sensitivity at 95\% C.L. for 30 days of counting
	at each mass splitting. The solid green and blue curves
	correspond to detected rates of $20$ and
	$1.24\times10^4\,\mathrm{s^{-1}}$, respectively, with
	$\beta_B=10^{-5}$ and fixed calibration. The dashed curves
	include $\beta_B=10^{-4}$, $\sigma_{\rm rel}=10^{-3}$,
	and $\sigma_{B_0}=2\,\mu\mathrm{T}$.
	The vertical dotted line indicates the assumed maximum
	magnetic field of $1\,\mathrm{T}$.
	The shaded regions correspond to existing laboratory
	constraints~\cite{Nesvizhevsky2006,Ban2007,Serebrov2009,BerezhianiNesti2012,Abel2021,Ban2023,StasserMURMUR2021,AlmazanSTEREO2022,Broussard2022,GonzalezSNS2024,Brenot2026,AyresMass2026}.
	(b) Median expected 95\% exclusion limits for the paired,
	three-field, and block acquisition protocols with 30 days
	of scheduled beam time. The solid and dashed curves
	correspond to the lower and higher detected rates,
	respectively. The gray region represents the exclusion
	limits from SNS and STEREO for positive mass
	splittings~\cite{GonzalezSNS2024}, while the green region
	indicates the additional parameter space accessible through
	the block acquisition with $K=512$ at $20\,\mathrm{s^{-1}}$.}
\label{fig:updated_vcn_reach}
\end{figure*}

Figure~\ref{fig:updated_vcn_reach}(a) shows the local projections
obtained by assigning the full on/off exposure to a single tested
splitting at a time. Outside $2$--$10\,\mathrm{neV}$ we retain the
same instrumental parameters and illustrate a maximum nominal field
of $1\,\mathrm{T}$ for both search and reference settings; the
operating range of a realized apparatus depends on its magnet geometry,
field uniformity, and magnetic isolation. Panel (b) instead assigns
a single beam allocation to the entire scan, as described in
Sec.~\ref{sec:scan_optimization}.

\section{Sensitivity at fixed total scan exposure}
	\label{sec:scan}

	The narrow magnetic resonance improves selectivity in mass splitting,
but an unknown resonance position requires distributing the exposure
among many field settings. We first quantify this statistical cost
with independent search/reference pairs and a fixed total live time,
using the reduced response of Eq.~\eqref{eq:active_response}. This
provides a baseline for the comparison at fixed scheduled beam time
in Sec.~\ref{sec:scan_optimization}.

The resonance condition maps $2$--$10\,\mathrm{neV}$ onto the field
interval from $(2\,\mathrm{neV})/\mun$ to $(10\,\mathrm{neV})/\mun$.
We extend both endpoints by $300\,\mu\mathrm{T}$ to sample the
resonance profile near the boundaries. Let $B_j$, $j=1,\ldots,N_B$,
be the nominal search fields and $h_B$ the grid spacing. Dividing
the extended field range by $h_B$ and rounding the number of intervals
upward gives the number of search settings,
	\begin{equation}
		N_B=1+\left\lceil\frac{(8\,\mathrm{neV})/\mun+600\,\mu\mathrm{T}}{h_B}\right\rceil.
		\label{eq:scan_number}
	\end{equation}
	Here $\lceil x\rceil$ denotes the smallest integer greater than or equal to $x$, and the additional unit counts the first grid point. Each search field is paired with a reference field displaced from it by at least $100\,\mu\mathrm{T}$ or eight effective rms field widths, whichever is larger,
	\begin{equation}
		B_j^{\rm ref}=B_j+\max(100\,\mu\mathrm{T},8\beta_B B_j).
	\end{equation}
	Reference exposures are counted separately even if their nominal fields coincide with other search settings. Each of the $2N_B$ acquisitions receives a live time $T_{\rm scan}/(2N_B)$, where $T_{\rm scan}=30$ days is the total counting exposure over the entire grid. The expected count per acquisition in the absence of mixing is $n_{\rm scan}=RT_{\rm scan}/(2N_B)$, with $R$ the reference detected rate. For $h_B=10\,\mu\mathrm{T}$, one obtains $N_B=13326$ pairs, corresponding to approximately $97\,\mathrm{s}$ of live time per acquisition.

	Unlike a locally tuned measurement, a scan cannot guarantee that every reference field is off resonance for the true splitting. Both measurements in each pair must therefore enter the signal prediction. Denoting the reduced survival response at an arbitrary nominal field $B$ by $g(B;\tau_{nn'},\dm,\delta B_0)$, we define the predicted ratio for pair $j$ as
	\begin{equation}
		r_j=(1+\eta_j)\frac{g(B_j;\tau_{nn'},\dm,\delta B_0)}
		{g(B_j^{\rm ref};\tau_{nn'},\dm,\delta B_0)}.
	\end{equation}
	Here $r_j$ is the dimensionless search/reference rate ratio and $\eta_j$ is the fractional relative response mismatch of pair $j$. Each pair has a free common rate normalization to allow for changes in the incident flux between pairs; the field-scale offset $\delta B_0$ is common to the full scan. Profiling the Poisson likelihood over these pair normalizations, as in Eq.~\eqref{eq:updated_poisson_ratio}, gives
	\begin{equation}
		\begin{split}
			Q_A^{\rm scan}=\min_{\{\eta_j>-1\},\delta B_0}\Bigg\{
			&2n_{\rm scan}\sum_{j=1}^{N_B}\ln\frac{(1+r_j)^2}{4r_j}\\
			&+\bm\eta^{\mathsf T}\mathsf C_{\eta}^{-1}\bm\eta
			+\frac{\delta B_0^2}{\sigma_{B_0}^2}\Bigg\},
		\end{split}
		\label{eq:scan_exact}
	\end{equation}
	where $\mathsf C_\eta$ is the covariance matrix of the residual relative calibration shifts (the diagonal entries are their variances and the off-diagonal entries encode correlations) and $\bm\eta=(\eta_1,\ldots,\eta_{N_B})^{\mathsf T}$. A vanishing uncertainty is understood to fix the associated parameter. For the baseline estimates in Table~\ref{tab:scan_reach} we use $\mathsf C_\eta=\sigma_{\rm rel}^2\mathsf{1}$, with independently constrained pair mismatches. Section~\ref{sec:scan_optimization} compares acquisition schedules with a shared physical model of the rate variations.

	To evaluate the scan response, it is useful to expand the two-segment survival probability as $g(B)=1-A(B;\dm,\delta B_0)/\tau_{nn'}^2+\mathcal O(\tau_{nn'}^{-4})$. The coefficient, with dimensions of time squared, is
	\begin{equation}
		A(B)=2\left\langle t(k)^2\sum_{s=\pm}w_s
		\operatorname{sinc}^2\left(\frac{(s\mun B'-\dm)t(k)}{2\hbar}\right)\right\rangle,
		\label{eq:scan_A}
	\end{equation}
	where $s=+1,-1$ labels the two spin branches, $w_+=P_0$, $w_-=1-P_0$, and the brackets denote the spectral and effective-field averages at nominal field $B$, with $B'$ distributed according to Eq.~\eqref{eq:field_kernel}. Define $a_j=A(B_j)-A(B_j^{\rm ref})$ and $\bm a=(a_1,\ldots,a_{N_B})^{\mathsf T}$. The leading fractional deficit in each pair is $a_j/\tau_{nn'}^2$, with its sign determined by which field yields the larger conversion response. The linearized ratio is $r_j-1\simeq\eta_j-a_j/\tau_{nn'}^2$. Thus the counting variance and the covariance of the relative response shifts combine as in the local calculation. Expanding Eq.~\eqref{eq:scan_exact} and profiling over $\bm\eta$ gives
	\begin{equation}
		\begin{split}
			Q_A^{\rm scan}\simeq\min_{\delta B_0}\left[
			\frac{\bm a^{\mathsf T}\mathsf V^{-1}\bm a}{\tau_{nn'}^4}
			+\frac{\delta B_0^2}{\sigma_{B_0}^2}\right],\\
			\mathsf V=\frac{2}{n_{\rm scan}}\mathsf{1}+\mathsf C_\eta .
		\end{split}
		\label{eq:scan_weak}
	\end{equation}
	The matrix $\mathsf{1}$ is the identity on the space of acquisition pairs and $\mathsf V$ is the total covariance of their fractional rate differences, including the Poisson contribution $2/n_{\rm scan}$ per pair. For a fixed absolute field scale ($\delta B_0=0$) and independent relative response shifts, the projected pointwise sensitivity is therefore
	\begin{equation}
		\tau_{95}^{\rm scan}\simeq
		\left[\frac{\sum_j a_j^2}{2.71[4N_B/(RT_{\rm scan})+\sigma_{\rm rel}^2]}\right]^{1/4}.
		\label{eq:scan_tau}
	\end{equation}
	Several pairs contribute to the same tested resonance through their respective coefficients $a_j$. In the fine-grid, statistics-dominated limit, $\sum_j a_j^2\simeq h_B^{-1}\int\dd B\,a(B)^2$, whereas $N_B$ also scales as $h_B^{-1}$. The leading sensitivity thus becomes approximately independent of the grid spacing once the response is well resolved, provided that the additional switching overhead is negligible. A coarser grid concentrates the exposure at fewer fields but can miss the center of a narrow resonance.

	The choice of grid spacing follows from the finite resonance width. Near $\dm=5\,\mathrm{neV}$, at $R=20\,\mathrm{s^{-1}}$ with $\beta_B=10^{-5}$ and fixed calibration, we sample 16 equally spaced resonance positions within one grid interval. Both $h_B=5$ and $10\,\mu\mathrm{T}$ give $\tau_{95}^{\rm scan}\simeq0.0246\,\mathrm{s}$, whereas spacings of $20$ and $40\,\mu\mathrm{T}$ give sensitivities between $0.0219$ and $0.0266\,\mathrm{s}$ depending on the resonance position. The similar ranges for the two coarser grids follow from the $100\,\mu\mathrm{T}$ reference-field displacement, which interleaves search and reference fields for the latter spacing. We adopt $h_B=10\,\mu\mathrm{T}$, which resolves the resonance without doubling the number of settings.
	\begin{table*}[t]
		\caption{Projected pointwise 95\% sensitivities in the paired scan with $h_B=10\,\mu\mathrm{T}$ and 30 days of total counting exposure. The entries are the smallest sensitivities among eight equally spaced relative positions of the resonance within one grid interval at each representative splitting. Fixed calibration uses $\beta_B=10^{-5}$; the second scenario uses $\beta_B=10^{-4}$, independent relative response uncertainties $\sigma_{\rm rel}=10^{-3}$ per pair, and a common field offset uncertainty $\sigma_{B_0}=2\,\mu\mathrm{T}$. Values in seconds are calculated using Eq.~\eqref{eq:scan_weak}.\label{tab:scan_reach}}
		\begin{ruledtabular}
			\begin{tabular}{c cc cc}
				& \multicolumn{2}{c}{$R=20\,\mathrm{s^{-1}}$} & \multicolumn{2}{c}{$R=1.24\times10^4\,\mathrm{s^{-1}}$}\\
				$\dm$ [neV] & Fixed calibration & Calibration included & Fixed calibration & Calibration included\\\hline
				2 & 0.0246 & 0.0241 & 0.1229 & 0.1070\\
				5 & 0.0246 & 0.0224 & 0.1227 & 0.0996\\
				10 & 0.0245 & 0.0201 & 0.1222 & 0.0891\\
			\end{tabular}
		\end{ruledtabular}
	\end{table*}

	The weak-mixing response was evaluated using Gaussian quadratures and interpolation in field detuning; doubling the quadrature order from 64 to 128, halving the interpolation spacing from $0.25$ to $0.125\,\mu\mathrm{T}$, and increasing the retained detuning window from 1 to 2 mT (with additional margin for the reference fields) changed the reported sensitivities by less than $10^{-6}$ in relative terms. The opposite spin branch is far from resonance over this positive-field interval; its contribution to $A$ is bounded by $2(1-P_0)(2\hbar/|\Delta_-|)^2\lesssim 3\times10^{-15}\,\mathrm{s^2}$ and was neglected in the scan quadrature. At the tabulated $10\,\mu\mathrm{T}$ sensitivities, recalculating the finite-mixing survival probabilities and profiling the Poisson statistic in Eq.~\eqref{eq:scan_exact} gives $Q_A^{\rm scan}$ within 2\% of 2.71, corresponding to approximately percent-level accuracy in the inferred oscillation time.

	With fixed calibration, the scan sensitivity at the reference rate is about one eighth of the corresponding local reach. At the higher rate, the corresponding scan sensitivity is approximately $0.12\,\mathrm{s}$, compared with the locally tuned target of $1\,\mathrm{s}$. With the assumed calibration uncertainties, effective magnetic-field broadening reduces the reach toward larger splittings. In contrast to a locally tuned measurement, the per-pair statistical variance $4N_B/(RT_{\rm scan})$ is larger than or comparable to $\sigma_{\rm rel}^2$ even in the higher-rate scan, so the scan remains substantially sensitive to an increase in neutron rate. The cost of dedicated reference exposures can be assessed by considering an alternative acquisition in which the flux and detector response are calibrated over the entire scan. In the fixed-calibration limit, a single exposure of duration $T_{\rm scan}/N_B$ is then assigned to each $B_j$, with one common free normalization. The weak-mixing Poisson statistic after profiling this normalization is
	\begin{equation}
		q_A^{\rm shared}\simeq\frac{RT_{\rm scan}}{N_B\tau_{nn'}^4}
		\sum_{j=1}^{N_B}\left[A(B_j)-\bar A\right]^2,
		\label{eq:scan_shared}
	\end{equation}
	where $\bar A=N_B^{-1}\sum_j A(B_j)$. At $h_B=10\,\mu\mathrm{T}$ and $\beta_B=10^{-5}$, this gives sensitivities of approximately $0.029\,\mathrm{s}$ and $0.146\,\mathrm{s}$ at the two rates considered in Table~\ref{tab:scan_reach}, respectively. The improvement of about 19\% relative to the fixed-calibration paired acquisition arises because all the exposure is spent on search settings and the normalization is constrained by the field sequence as a whole. Achieving that improvement would require relative stability across the scan, whereas the paired protocol allows changes in the common flux between pairs.

	\subsection{Common instrumental model at fixed beam time}
\label{sec:scan_optimization}

The preceding calculation treats the calibration uncertainty of each
pair independently. To compare different acquisition protocols, we
instead describe the same sources of rate variation in all schedules,
allowing their correlations to depend on the actual fields and times
at which the counts are acquired. We fix the scheduled beam time
$T_{\rm sched}=30$ days, including intervals without counting.
Index $i$ now labels individual acquisitions in chronological order,
with nominal field $B_i$, live time $T_i$, and start time $t_i$.
For summed counts $N_i$, define the dimensionless fractional rate
$y_i=N_i/(RT_i)-1$. To first order in mixing and instrumental rate
variations, we write
\begin{equation}
 y_i=\nu_{g(i)}-\frac{A(B_i)}{\tau_{nn'}^2}
     +\bar f_i+s(B_i)+u_i+e_i ,
 \label{eq:common_scan_rate}
\end{equation}
where $A(B_i)$ is given by Eq.~\eqref{eq:scan_A} and the free
fractional normalization $\nu_{g(i)}$ is common to acquisitions in
group $g(i)$ (a pair, triplet, or block). The remaining terms describe
different physical sources of fluctuations. The drift $f(t)$ accounts
for temporal changes of the incident flux and beam transport, and
$\bar f_i=T_i^{-1}\int_{t_i}^{t_i+T_i} f(t)\,\dd t$ is its average
during acquisition $i$. The function $s(B)$ represents a fractional
transmission variation reproducible at a given magnetic field, while
$u_i$ represents independent repeatability fluctuations between
acquisitions. Counting noise and the residual uncertainty from the
port correction of Sec.~\ref{sec:port_budget} are included in $e_i$.

We model $f$ and $s$ as independent zero-mean Gaussian processes.
Their rms amplitudes are $\sigma_f$ and $\sigma_s$, respectively;
$t_c$ is the time difference and $B_c$ the field difference at which
the corresponding correlation decreases to $e^{-1}$ of its zero-
separation value. Averaging the temporal covariance over both counting
intervals, but evaluating the transmission at the nominal fields,
gives the contributions
\begin{align}
 C^{(t)}_{ij}
 &=\frac{\sigma_f^2}{T_iT_j}
   \int_{t_i}^{t_i+T_i}\!\dd t
   \int_{t_j}^{t_j+T_j}\!\dd t'\,
      e^{-|t-t'|/t_c}, \label{eq:common_time_cov}\\
 C^{(B)}_{ij}
 &=\sigma_s^2 e^{-|B_i-B_j|/B_c}.
 \label{eq:common_field_cov}
\end{align}
The full covariance of the fractional rates is
$C_{ij}=C^{(t)}_{ij}+C^{(B)}_{ij}
+\delta_{ij}[1/(RT_i)+\sigma_u^2]+C^{\rm port}_{ij}$,
where $\delta_{ij}$ is the Kronecker delta, $\sigma_u$ is the rms
repeatability error, and $C^{\rm port}_{ij}$ is the residual
port-correction covariance specified in Sec.~\ref{sec:port_budget}.
The term $1/(RT_i)$ follows from the Poisson variance of $N_i$ in
the no-mixing limit. Correlations between different groups are retained
in this covariance. For the reference model we adopt
$(\sigma_f,t_c)=(10^{-3},2\,\mathrm{h})$,
$(\sigma_s,B_c)=(5\times10^{-4},50\,\mu\mathrm{T})$, and
$\sigma_u=2\times10^{-4}$. These illustrative rate variations are
independent of the magnetic-field distribution sampled by the neutrons,
for which we use $\beta_B=10^{-4}$ in Eq.~\eqref{eq:field_kernel}.

We retain the $10\,\mu\mathrm{T}$ search grid, its endpoint margins,
and the field displacement
$\Delta_j=\max(100\,\mu\mathrm{T},8\beta_B B_j)$.
A pair estimates the local normalization from a detuned acquisition;
a triplet brackets the search setting in time and field to reduce
sensitivity to slow drift. A block instead uses the fields away from
the resonance within the same group to constrain the normalization.
We compare the following sequences, traversing the search grid in
increasing field order.
\begin{enumerate}
\item A pair uses $B_j$ and $B_j+\Delta_j$ for equal live times.
\item A triplet uses $B_j-\Delta_j$, $B_j$, and $B_j+\Delta_j$ with
live-time fractions $1/4$, $1/2$, and $1/4$. This ordering follows
the bracketing principle used in the UCN search at
PF2~\cite{Brenot2026}, with detunings chosen for the VCN line shape.
\item A block shares a normalization among consecutive search
settings. Each block is contained within one exposure band, and we
take maximum block sizes $K=32,128,512$, without dedicated reference
acquisitions.
\end{enumerate}
For a field change $\Delta B_i=B_i-B_{i-1}$, we exclude
$t_0+|\Delta B_i|/v_B$ from live time, with
$t_0=5\,\mathrm{s}$ and $v_B=100\,\mu\mathrm{T\,s^{-1}}$ as design
inputs. Including transitions between groups, the live time is
$T_{\rm live}=T_{\rm sched}-\sum_{i>1}(t_0+|\Delta B_i|/v_B)$,
with the first acquisition starting at its prescribed field.
The starting times obey $t_i=t_{i-1}+T_{i-1}+t_0+|\Delta B_i|/v_B$
for $i>1$. The schedules reported below assume that auxiliary
calibrations are available before the 30-day beam allocation. If a
calibration of duration $T_{\rm cal}$ is performed within this
allocation, its duration and any additional switching overhead must
be subtracted from $T_{\rm live}$; calibrations interleaved with
counting also modify the acquisition clock entering
Eq.~\eqref{eq:common_time_cov}. The rate-dependent requirements for
the port-efficiency calibration are examined in Sec.~\ref{sec:port_budget}.

Since each group has a free normalization, the signal is constrained
by the relative rates within that group rather than by their absolute
levels. To remove the normalization, define the matrix $\mathsf D$
with one row per consecutive pair of acquisitions within a group,
containing $-1$ and $+1$ at the corresponding positions and zero
elsewhere. Applied to the rate vector $\bm y$, this yields the
independent linear contrasts $\mathsf D\bm y$, whose covariance is
$\mathsf D\mathsf C\mathsf D^{\mathsf T}$. A group of $m$ acquisitions
contributes $m-1$ linearly independent contrasts, which need not be
statistically independent. In the Gaussian approximation these
contrasts retain the information after profiling the group
normalizations. Writing $A_i=A(B_i)$, $\bm A=(A_1,A_2,\ldots)^{\mathsf T}$
and $\bm a=\mathsf D\bm A$, the expected contrast in the mixing
hypothesis is $-\bm a/\tau_{nn'}^2$. The Asimov statistic is therefore
\begin{equation}
 Q_A(\tau_{nn'},\dm)\simeq
  \frac{\bm a^{\mathsf T}
       (\mathsf D\mathsf C\mathsf D^{\mathsf T})^{-1}\bm a}
       {\tau_{nn'}^4}.
 \label{eq:shared_schedule_information}
\end{equation}
At fixed field offset the projected sensitivity is
$\tau_{95}=[\mathcal F(\dm)/2.71]^{1/4}$, where
$\mathcal F=\bm a^{\mathsf T}(\mathsf D\mathsf C\mathsf D^{\mathsf T})^{-1}\bm a$
has dimensions of time to the fourth power. Both side-field conversion
responses enter $\bm A$ in a triplet. Using only the central rate
minus the average of its side rates, $y_{\rm center}-(y_{\rm left}+y_{\rm right})/2$,
reproduces the linearized $R_{ABC}$ statistic of Ref.~\cite{Brenot2026}.
For a locally linear temporal drift, that combination cancels the
slope at equal intervals between the acquisition midpoints.

The unknown splitting also determines how the exposure should be
allocated: the aim is to improve the weakest sensitivity in the
entire mass interval. We divide the grid into four consecutive bands
$r=1,\ldots,4$, with approximately equal numbers $M_r$ of search
settings. Let $w_r$ be the fraction of $T_{\rm live}$ assigned to
band $r$, with $w_r\geq0$ and $\sum_r w_r=1$. For an acquisition $i$
in that band, the live time is $T_i=c_i w_r T_{\rm live}/M_r$,
where $c_i=1/2$ for either member of a pair, $c_i=1/4,1/2,1/4$
for the successive members of a triplet, and $c_i=1$ for a search
setting in a block. Each trial allocation determines both the counting
variances and the temporal covariance through the acquisition clock.

For each schedule we select the allocation that maximizes
$\min_{\dm\in[2,10]\,\mathrm{neV}}\tau_{95}(\dm)$ among the tested
allocations. Starting from uniform exposure and from allocations
weighted by the inverse information at the least sensitive points of
each band, we transfer time toward the limiting bands and reevaluate
the minimum on a mass grid corresponding to $2.5\,\mu\mathrm{T}$ in
field, refined near the minima. Table~\ref{tab:optimized_scan} reports
the resulting minimum pointwise sensitivities.

\begin{table*}[t]
 \caption{Minimum pointwise $\tau_{95}$ over the positive branch of
 $2$--$10\,\mathrm{neV}$ for the common rate model and a 30-day
 beam allocation, with auxiliary calibrations available beforehand.
 The quoted live times account for field changes. Each protocol has
 its own optimized band fractions.
 \label{tab:optimized_scan}}
 \begin{ruledtabular}
 \begin{tabular}{lrrr}
 Protocol & Live time (days) & $\tau_{95}^{\min}$ (s), $R=20\,\mathrm{s^{-1}}$
  & $\tau_{95}^{\min}$ (s), $R=1.24\times10^4\,\mathrm{s^{-1}}$\\
 \hline
 Pairs & 28.15 & 0.0210 & 0.0853 \\
 Triplets, both contrasts & 27.05 & 0.0220 & 0.0883 \\
 Blocks, $K=32$ & 29.21 & 0.0239 & 0.0759 \\
 Blocks, $K=128$ & 29.21 & 0.0248 & 0.0775 \\
 Blocks, $K=512$ & 29.21 & 0.0250 & 0.0805
 \end{tabular}
 \end{ruledtabular}
\end{table*}

At the lower rate, counting statistics favor blocks, which allocate
the entire live time to search settings and share a normalization.
The $K=128$ case captures almost all the gain of the larger blocks
while requiring stable operation over a shorter interval. At the higher
rate, the increased importance of instrumental fluctuations favors the
two-contrast triplet in the reference model. Using only the central-minus-side
combination on the same triplet schedule gives minimum times of
$0.0193$ and $0.0786\,\mathrm{s}$ at the two rates. Conversion in a side
setting and the change in the rate between those settings provide
additional information in the full fit.

To determine how the choice of protocol depends on the underlying
instrumental response, we vary the same covariance parameters for
all schedules, leaving the other reference inputs fixed. At the higher rate, increasing
$\sigma_s$ to $10^{-3}$ and reducing $B_c$ to $20\,\mu\mathrm{T}$
gives $0.0699\,\mathrm{s}$ for triplets. A faster drift with
$(\sigma_f,t_c)=(2\times10^{-3},10\,\mathrm{min})$ gives
$0.0851\,\mathrm{s}$ for triplets and $0.0551\,\mathrm{s}$ for
$K=512$ blocks. For slower, smaller drift
$(\sigma_f,t_c)=(3\times10^{-4},8\,\mathrm{h})$ and smoother
transmission $(\sigma_s,B_c)=(2\times10^{-4},250\,\mu\mathrm{T})$,
the $K=512$ blocks reach $0.1123\,\mathrm{s}$ and the
triplets $0.1062\,\mathrm{s}$. Varying $t_0$ from $2$ to
$20\,\mathrm{s}$ in the reference model leaves the ordering of the
best protocols unchanged. The longer settling time reduces the
triplet live time from $28.44$ to $20.11$ days.

At the reference-model minima, evaluating the finite-mixing
conversion in the resonant spin branch changes the time limits by less than $1\%$. Enlarging the
retained field window around each representative minimum changes
the sensitivities by less than $0.2\%$. A common field offset within
$\sqrt{2.71}\,\sigma_{B_0}$ translates the mass argument in the
resonant spin branch. Extending the checked mass range accordingly
changes the minimum by less than $0.02\%$ in the reference schedules.
As a counting-level check we generated 1200 joint realizations of
the drift and transmission processes for each rate, evaluating the
same realizations on paired, triplet, and $K=128$ schedules.
Poisson counts were drawn in each realization with and without
a finite-mixing signal at $\dm=8\,\mathrm{neV}$. The fitted signal strength, proportional to $\tau_{nn'}^{-2}$,
was compared with its injected value. The residuals divided by the
forecast standard error had widths between 0.97 and 1.03, testing
the predicted statistical dispersion at that splitting.

\subsection{Projected extension of laboratory exclusion}
\label{sec:scan_exclusion}

The relevant phenomenological question is whether the sensitivity
of a complete scan exceeds the existing laboratory bounds at the
same splitting. In Fig.~\ref{fig:updated_vcn_reach}(b), let
$\tau_{\rm lab}(\dm)$ be the larger of the SNS and recast STEREO
95\% lower bounds on the oscillation time~\cite{AlmazanSTEREO2022,GonzalezSNS2024}. The green band
runs from $\tau_{\rm lab}$ up to the median expected limit
$\tau_{95}^{\rm scan}$ for the $K=512$ block acquisition at
$20\,\mathrm{s^{-1}}$. In mixing energy the corresponding interval is
\begin{equation}
 \frac{\hbar}{\tau_{95}^{\rm scan}(\dm)}
 \ \lesssim\ \epsnn\ <\ \frac{\hbar}{\tau_{\rm lab}(\dm)} .
 \label{eq:projected_exclusion_band}
\end{equation}
This interval describes the extension expected in a null scan
under the nominal response model and on the branch connected
to weak mixing. The smallest ratio
$\tau_{95}^{\rm scan}/\tau_{\rm lab}$ is about two, near
$\dm=10\,\mathrm{neV}$. There the reconstructed STEREO bound is
about $13\,\mathrm{ms}$ and the VCN projection is
$25\,\mathrm{ms}$. Taking the intersection of Eq.~\eqref{eq:projected_exclusion_band}
over the entire positive-splitting interval gives a conservative
common mixing-energy range of approximately
$(2.8\text{--}4.8)\times10^{-14}\,\mathrm{eV}$.
These comparisons use laboratory data. Bounds based on neutron-star
temperatures also depend on the conversion and thermal evolution of
dense matter in the assumed hidden-sector model~\cite{McKeenHeating2021,AyresMass2026}.

\section{Accepted neutron rate and calibration of the two outputs}
\label{sec:high_rate}

\subsection{Accepted phase space}

The reference rate must be related to the number of neutrons that
can be transported through the interferometer. We estimate this
rate for a collimated $40\,\text{\AA}$ beam using the geometry of
Fig.~\ref{fig:schema_v5}. The incident spectral brightness
$\mathcal B_{\rm in}(\lambda,\Omega)$ is the neutron rate per unit
entrance area, wavelength interval, and solid angle. Integrating
over the accepted phase space gives
\begin{equation}
 R_{\rm det}=A\int\dd\lambda\int\dd\Omega\,
   \mathcal B_{\rm in}(\lambda,\Omega)
   \mathcal T(\lambda,\Omega)\eta_{\rm det}(\lambda,\Omega).
 \label{eq:transport_integral}
\end{equation}
Here $A$ is the entrance area, $\mathcal T$ includes the selection and
transport through the optics, and $\eta_{\rm det}$ is the mean
detection efficiency. We take two aligned $3\times15\,\mathrm{mm^2}$
slits $0.60\,\mathrm m$ apart. The slit size is taken from the
PF2 VCN geometry of Ref.~\cite{Oda2017}. In the paraxial
approximation and for uniform illumination in the accepted phase
space, the area is $A=0.45\,\mathrm{cm^2}$ and the geometrical solid
angle is $\Omega_{\rm eff}=1.25\times10^{-4}\,\mathrm{sr}$. The
effective solid angle follows from the overlap of two identical
rectangular slits. For slit width $w$, height $h$, and separation
$\ell$, the transmitted fraction at small angles $\theta_x,\theta_y$
is $(1-\ell|\theta_x|/w)_+(1-\ell|\theta_y|/h)_+$, where
$(z)_+=\max(z,0)$. Its angular integral is $wh/\ell^2$;
this accounts for the decreasing overlap at nonzero incidence angles.

We model the wavelength selector by a unit-peak transmission
proportional to $(k/k_0)^2
\exp[-(k-k_0)^2/(2\sigma_k^2)]$, with $\sigma_k/k_0=0.10$.
For a locally flat incident spectral brightness, the factor
$(k/k_0)^2$ compensates the wavelength-to-wave-number Jacobian and
yields the Gaussian accepted spectrum used in
Eq.~\eqref{eq:spectral_average}. The normalization is fixed by the maximum of the transmission
profile, so its wavelength integral is $9.83\,\text{\AA}$. We propagate rays through the two
slits and along the horizontal interferometer with gravity included,
centering the downstream apertures on the $40\,\text{\AA}$
trajectory. We place the downstream slit at the beam splitter
and evaluate clear apertures at the central mirrors, recombiner,
and detectors at path distances of $0.50$, $1.00$, and $1.50\,\mathrm{m}$.
Each aperture has a projected full width of $20\,\mathrm{mm}$
and a full height of $80\,\mathrm{mm}$ in the transport calculation.
We apply the nickel critical-angle condition at a
$3^\circ$ central grazing angle (measured from the reflecting
surface), allowing for the divergence set by the slits~\cite{OriginalProposal,Schanzer2016}.
The aperture and reflection-angle cuts together retain about $98\%$
of the already collimated and wavelength-selected rays. The mean
squared flight time in an active segment changes by about $0.15\%$
under these cuts.

For the transport estimate we take an additional guide transmission
of $0.92$, a transmission of $0.44$ for polarization selection from
an unpolarized beam, a total reflected-plus-transmitted fraction of
$0.98$ at each of the beam splitter and recombiner, and a reflectivity
of $0.995$ for each central mirror. Each detected ray encounters one
central mirror. We use $0.995$ for the phase-shifter transmission in
arm II and $0.85$ for the mean detector efficiency. These design
inputs give a net transmission times detection efficiency of about
$0.32$ after the slit and spectral factors, including the geometric
cuts. Approximating the incident brightness as constant across the accepted
wavelength and angular ranges, the rate follows by multiplying the
entrance area, the effective solid angle, the selector integral, and
the net transmission times detection efficiency. This gives
\begin{equation}
 R_{\rm det}\simeq(1.78\times10^{-4}\,
    \mathrm{cm^2\,\text{\AA}\,sr})\,\mathcal B_{40},
 \label{eq:rate_effective_acceptance}
\end{equation}
where $\mathcal B_{40}$ denotes the incident spectral brightness near
$40\,\text{\AA}$. The optical transmissions, illuminated
footprints, and local spectral brightness can be replaced by measured
quantities for a particular realization.

Using the effective PF2 brightness scale
$(1\text{--}2)\times10^5\,\mathrm{s^{-1}cm^{-2}\,\text{\AA}^{-1}sr^{-1}}$
from Ref.~\cite{OriginalProposal} gives $18$--$36\,\mathrm{s^{-1}}$
for this acceptance. The PF2 flux-density measurement in
Ref.~\cite{Neulinger2024} refers to a different collimated beam
and provides an independent scale check. At fixed geometry the
higher-rate target of $1.24\times10^4\,\mathrm{s^{-1}}$ requires
an average brightness delivered to the instrument of about
$6.9\times10^7\,\mathrm{s^{-1}cm^{-2}\,\text{\AA}^{-1}sr^{-1}}$.
This requirement can also be expressed as a target on the product
of accepted phase space and transmission for a proposed
long-wavelength beamline~\cite{Andersen-2020,Zimmer,HighNESSCDR2024}.
The relevant rate is the detected mean over live time, including any
time-of-flight acceptance.

\subsection{Impact of correcting port redistribution}
\label{sec:port_budget}

When the two detector efficiencies differ, a field-dependent change
of the interference phase can mimic disappearance through
Eq.~\eqref{eq:port_spurious}. Since both output intensities are
measured, their efficiency-corrected sum is
$\overline I_{+,i}=J_{H,i}/e_H+J_{O,i}/e_O$. Expressing the two
terms in terms of $J_{+,i}$ and the measured asymmetry
$C_i^{\rm det}$ gives
\begin{equation}
 \overline I_{+,i}
   =\frac{J_{+,i}(1-d_{\rm e}C_i^{\rm det})}
          {\bar e(1-d_{\rm e}^2)} .
 \label{eq:port_corrected_intensity}
\end{equation}
The common denominator cancels in an on/off ratio. An independent
calibration of $d_{\rm e}$ allows the port counts acquired with
the signal sample to determine the field-correlated change in
asymmetry. We take a stable $d_{\rm e}=0.01$ and a calibration
uncertainty $\sigma_{d_{\rm e}}=10^{-4}$. At the assumed mean
efficiency $0.85$ the two efficiencies are 0.8585 and 0.8415.
With equal known illumination of the two ports, $d_{\rm e}$ can
be estimated from the asymmetry of the calibration counts. For
$N_{\rm cal}$ detected calibration events, the conditional binomial
variance is $(1-d_{\rm e}^2)/N_{\rm cal}$. Thus, at a total
calibration rate $R_{\rm cal}$, the counting time required to achieve
a precision $\sigma_{d_{\rm e}}$ is approximately
$T_{\rm cal}=1/(R_{\rm cal}\sigma_{d_{\rm e}}^2)$ for
$|d_{\rm e}|\ll1$. Reaching $10^{-4}$ requires about $10^8$
detected events, or $2.2\,\mathrm{h}$ at
$R_{\rm cal}=1.24\times10^4\,\mathrm{s^{-1}}$, but approximately
58 days at $R_{\rm cal}=20\,\mathrm{s^{-1}}$. The latter exceeds
the entire search allocation. At the PF2-scale rate, such precision
would therefore require an independent calibration, for example in
a higher-rate beam before the search, with the relative efficiency
of the installed detection channels verified under the operating
conditions. Alternatively, a precision of $10^{-3}$ requires only
$10^6$ events, corresponding to about 14 h at the lower rate.
These are counting-statistics requirements; the relative illumination
and the stability of the calibrated efficiencies must also be
controlled.

Let $x=C_{\rm on}-C_{\rm off}$. For a reference rms
$\sigma_{x,0}=0.02$ of uncorrected redistribution, the associated
fractional uncertainty in the summed-rate contrast is
$|d_{\rm e}|\sigma_{x,0}$. After the correction, the remaining
uncertainty arises from the statistical precision of the measured
asymmetry difference and the independent calibration of $d_{\rm e}$.
Linear propagation of these two uncertainties gives
\begin{equation}
 \sigma^2_{\rm port,corr}=
   d_{\rm e}^2\sigma_{x,\rm counts}^2+
   \sigma_{d_{\rm e}}^2\sigma_{x,0}^2 .
 \label{eq:port_variance_transport}
\end{equation}
Near balanced output intensities,
$\sigma_{x,\rm counts}\simeq2/\sqrt{RT_{\rm tot}}$ for a local
on/off pair with equal live times. We add the variance in
Eq.~\eqref{eq:port_variance_transport} to the summed-rate
counting variance and the remaining independent rate-response
variance $\sigma_\eta^2$ of Eq.~\eqref{eq:joint_port_profile}.
For Table~\ref{tab:port_reach} the total variance before correction is
$4/(RT_{\rm tot})+\sigma_\eta^2+d_{\rm e}^2\sigma_{x,0}^2$;
after correction the final term is replaced by
$\sigma^2_{\rm port,corr}$. The corresponding limits follow from
$Q_A\simeq A_2^2/(\tau_{nn'}^4\sigma_{\rm tot}^2)=2.71$, with
$\sigma_{\rm tot}^2$ the appropriate total variance.

\begin{table}[t]
 \caption{Local 95\% sensitivities in $\tau_{nn'}$ with the
 same summed rate before and after correcting port redistribution.
 The example uses $\dm=10\,\mathrm{neV}$,
 $\beta_B=10^{-5}$, a fixed field scale, and 30 days of live
 time, with an independently available efficiency calibration
 $\sigma_{d_{\rm e}}=10^{-4}$. The variable $\sigma_\eta$ gives the
 residual relative rate uncertainty from effects common to the two ports.
 \label{tab:port_reach}}
 \begin{ruledtabular}
 \begin{tabular}{rccc}
 $R\ (\mathrm{s^{-1}})$ & $\sigma_\eta$ &
 $\tau_{95}$ before (s) & $\tau_{95}$ after (s) \\
 \hline
 20 & $10^{-3}$ & 0.103 & 0.104 \\
 20 & $10^{-4}$ & 0.177 & 0.195 \\
 $1.24\times10^4$ & $10^{-5}$ & 0.236 & 0.860
 \end{tabular}
 \end{ruledtabular}
\end{table}

At the higher rate the port contribution falls from
$2.0\times10^{-4}$ to $2.0\times10^{-6}$ in this local example.
Table~\ref{tab:port_reach} shows how the resulting gain depends
on the precision of the remaining rate calibration. At the lower
rate, relaxing $\sigma_{d_{\rm e}}$ from $10^{-4}$ to $10^{-3}$
increases the calibration contribution to the fractional contrast
from $2\times10^{-6}$ to $2\times10^{-5}$, still below the counting
uncertainty of $2.78\times10^{-4}$. For $\sigma_\eta=10^{-3}$ and
$10^{-4}$, the corrected local sensitivities become approximately
$0.10376$ and $0.19433\,\mathrm{s}$, respectively. If the roughly
14 h required for this calibration at $20\,\mathrm{s^{-1}}$ are
taken from a 30-day allocation before the local measurement,
the remaining 29.42 days of counting yield approximately $0.10372$
and $0.19350\,\mathrm{s}$, neglecting additional calibration setup
time. The loss relative to the independently calibrated 30-day
projection is below $0.6\%$ in both cases. The corrections can be
applied within wavelength bins when port efficiencies or redistribution
depend on wavelength; the available calibration counts must then be
accounted for in each bin.

For the scheduled scans we take an rms per-setting asymmetry of
$0.02$ with a field correlation length of $50\,\mu\mathrm{T}$.
The corresponding residual port covariance in Eq.~\eqref{eq:shared_schedule_information}
is, near balanced ports,
$C^{\rm port}_{ij}=\delta_{ij}d_{\rm e}^2/(RT_i)
+\sigma_{d_{\rm e}}^2(0.02)^2\exp[-|B_i-B_j|/(50\,\mu\mathrm{T})]$.
Before correction, the redistribution covariance instead has amplitude
$d_{\rm e}^2(0.02)^2$ and the same field correlation. The latter is the design covariance averaged over
asymmetry profiles for a single common $d_{\rm e}$. For an
actual measured profile, its calibration term is proportional
to $C_i C_j$. With the reference scan noise and the same
exposure allocations, the correction changes the projected
minimum by less than $0.02\%$ at $20\,\mathrm{s^{-1}}$ and by
$1.1$--$2.2\%$ at the higher rate. This smaller gain reflects the
counting variance at each field and the other setting-dependent
transmission errors. In particular, the low-rate scan can be carried
out without the port-redistribution correction, retaining the same
exposure allocation and the uncorrected redistribution covariance
of this model. Its projected reach thus does not require the 58-day
efficiency calibration at the search rate. A drift common to both
ports still requires a separate rate constraint.

\section{Discussion and conclusions}
\label{sec:conclusions}

We have connected the resonant neutron--hidden-neutron conversion
response of a VCN Mach--Zehnder interferometer to the statistical
and instrumental requirements of a search at unknown mass splitting.
By comparing acquisition schedules under a common rate-covariance
model and at fixed scheduled beam time, we find that a 30-day block
scan with $20\,\mathrm{s^{-1}}$ at the detectors yields a minimum
median expected 95\% lower limit of approximately $25\,\mathrm{ms}$
over the positive-splitting interval $2$--$10\,\mathrm{neV}$.
Near $10\,\mathrm{neV}$ the projected limit exceeds the reconstructed
laboratory bound of about $13\,\mathrm{ms}$ by approximately a factor
of two. Under the laboratory comparison adopted in
Fig.~\ref{fig:updated_vcn_reach}(b), a null scan could extend the
exclusion across the entire interval to mixing energies in the range
$(2.8\text{--}4.8)\times10^{-14}\,\mathrm{eV}$.

The one-second sensitivity target refers instead to a run tuned to
a chosen splitting and requires a high detected rate and stringent
relative calibration. In a scan, sharing a normalization across
128 consecutive settings recovers nearly all the gain of longer
blocks at the lower rate. At the higher rate, a symmetric three-field
sequence gives the greatest minimum sensitivity in the reference
drift model, whereas smaller and slower drift and smoother
field-dependent transmission favor blocks, with a projected reach
of about $0.11\,\mathrm{s}$. The temporal and magnetic correlations
of the rate therefore determine the choice of acquisition protocol.

The adopted collimation and transport yield an estimated detected
rate of $18$--$36\,\mathrm{s^{-1}}$ at the reference PF2 brightness.
The higher-rate scenario requires greater incident brightness or
usable phase-space acceptance. The interferometric readout provides
an independent measurement of intensity redistribution: correcting
for unequal port efficiencies increases the locally tuned high-rate
sensitivity from $0.24$ to $0.86\,\mathrm{s}$ when the remaining
relative rate uncertainty is $10^{-5}$ and the port-efficiency
calibration is independently available at the $10^{-4}$ level.
At the PF2-scale rate, the scan reach is essentially unchanged
without the redistribution correction, avoiding the need for this
stringent calibration within the search allocation. Measurements of the accepted spectrum, the field
along neutron trajectories, the temporal and field covariance of the
rate, and the detector efficiencies would permit these projections
to be evaluated for a specific beamline and guide the choice of
scan sequence.

\section*{Acknowledgments}
	A. C. and G. P. acknowledge partial financial support from MIUR.
	The authors acknowledge support from the Istituto Nazionale di Fisica
	Nucleare (INFN) through the Iniziativa Specifica QGSKY.

\section*{Data Availability}

	The numerical data and computational codes supporting
	the findings of this study are available from the
	corresponding author upon reasonable request.


\begin{thebibliography}{99}

		\bibitem{DS5}
		G. Bertone and D. Hooper, Rev. Mod. Phys. \textbf{90}, 045002 (2018).

		\bibitem{MM1}
		I. Yu. Kobzarev, L. B. Okun, and I. Ya. Pomeranchuk, Sov. J. Nucl. Phys. \textbf{3}, 837 (1966).

		\bibitem{MM2}
		S. I. Blinnikov and M. Yu. Khlopov, Sov. J. Nucl. Phys. \textbf{36}, 472 (1982).

		\bibitem{MM4}
		H. M. Hodges, Phys. Rev. D \textbf{47}, 456 (1993).

		\bibitem{MM3}
		Z. Berezhiani, Int. J. Mod. Phys. A \textbf{19}, 3775 (2004).

		\bibitem{MM5}
		R. Foot, Int. J. Mod. Phys. A \textbf{29}, 1430013 (2014).

		\bibitem{BerezhianiBento2006}
		Z. Berezhiani and L. Bento, Phys. Rev. Lett. \textbf{96}, 081801 (2006).

		\bibitem{BerezhianiReview}
		Z. Berezhiani, Eur. Phys. J. C \textbf{64}, 421 (2009).

		\bibitem{Nesvizhevsky2006}
		V. V. Nesvizhevsky \textit{et al.}, Phys. Rev. Lett. \textbf{97}, 132502 (2006).

		\bibitem{Ban2007}
		G. Ban \textit{et al.}, Phys. Rev. Lett. \textbf{99}, 161603 (2007).

		\bibitem{Serebrov2009}
		A. Serebrov \textit{et al.}, Nucl. Instrum. Methods Phys. Res. A \textbf{611}, 263 (2009).

		\bibitem{BerezhianiNesti2012}
		Z. Berezhiani and F. Nesti, Eur. Phys. J. C \textbf{72}, 1974 (2012).

		\bibitem{Abel2021}
		C. Abel \textit{et al.}, Phys. Lett. B \textbf{812}, 135993 (2021).

		\bibitem{Ban2023}
		G. Ban \textit{et al.}, Phys. Rev. Lett. \textbf{131}, 191801 (2023).

		\bibitem{Brenot2026}
		G. Brenot \textit{et al.}, Phys. Lett. B \textbf{879}, 140673 (2026).

		\bibitem{Sarrazin2015}
		M. Sarrazin \textit{et al.}, Phys. Rev. D \textbf{91}, 075013 (2015).

		\bibitem{Sarrazin2016}
		M. Sarrazin \textit{et al.}, Phys. Lett. B \textbf{758}, 14 (2016).

		\bibitem{StasserMURMUR2021}
		C. Stasser \textit{et al.}, Eur. Phys. J. C \textbf{81}, 17 (2021).

		\bibitem{AlmazanSTEREO2022}
		H. Almaz\'an \textit{et al.} (STEREO Collaboration), Phys. Rev. Lett. \textbf{128}, 061801 (2022).

		\bibitem{Broussard2022}
		L. J. Broussard \textit{et al.}, Phys. Rev. Lett. \textbf{128}, 212503 (2022).

		\bibitem{GonzalezSNS2024}
		F. M. Gonzalez \textit{et al.}, Phys. Rev. D \textbf{110}, 072022 (2024).

		\bibitem{MM6}
		M. Hostert, D. McKeen, M. Pospelov, and N. Raj, Phys. Rev. D \textbf{107}, 075034 (2023).

		\bibitem{Ebisawa}
		T. Ebisawa \textit{et al.}, Nucl. Instrum. Methods Phys. Res. A \textbf{344}, 597 (1994).

		\bibitem{Zeilinger}
		G. van der Zouw, M. Weber, J. Felber, R. G\"ahler, P. Geltenbort, and A. Zeilinger, Nucl. Instrum. Methods Phys. Res. A \textbf{440}, 568 (2000).

		\bibitem{Fujiie}
		T. Fujiie \textit{et al.}, Phys. Rev. Lett. \textbf{132}, 023402 (2024).

		\bibitem{Ackermann2026}
		R. H. Ackermann \textit{et al.}, Proc. SPIE \textbf{14102}, 1410207 (2026).

		\bibitem{OriginalProposal}
		A. Capolupo, G. Pisacane, A. Quaranta, and P. B\"oni, Phys. Rev. D \textbf{114}, 095021 (2026).

		\bibitem{Neulinger2024}
		T. Neulinger \textit{et al.}, Nucl. Instrum. Methods Phys. Res. A \textbf{1066}, 169570 (2024).

		\bibitem{Oda2017}
		T. Oda \textit{et al.}, Nucl. Instrum. Methods Phys. Res. A \textbf{860}, 35 (2017).

		\bibitem{Lemmel2010}
		H. Lemmel and A. G. Wagh, Phys. Rev. A \textbf{82}, 033626 (2010).

		\bibitem{MuPAD}
		M. Janoschek, S. Klimko, R. G\"ahler, B. Roessli, and P. B\"oni, Physica B \textbf{397}, 125 (2007).

		\bibitem{Cowan2011}
		G. Cowan, K. Cranmer, E. Gross, and O. Vitells, Eur. Phys. J. C \textbf{71}, 1554 (2011).

		\bibitem{AyresMass2026}
		N. J. Ayres \textit{et al.}, Phys. Rev. D \textbf{114}, 072015 (2026).

		\bibitem{Andersen-2020}
		K. H. Andersen \textit{et al.}, Nucl. Instrum. Methods Phys. Res. A \textbf{957}, 163402 (2020).

		\bibitem{Zimmer}
		O. Zimmer \textit{et al.}, J. Neutron Res. \textbf{24}, 95 (2022).

		\bibitem{HighNESSCDR2024}
		V. Santoro \textit{et al.}, J. Neutron Res. \textbf{25}, 85 (2024).

		\bibitem{Schanzer2016}
		C. Schanzer, M. Schneider, and P. B\"oni, J. Phys.: Conf. Ser. \textbf{746}, 012024 (2016).

		\bibitem{Kolevatov2019}
		R. Kolevatov, C. Schanzer, and P. B\"oni, Nucl. Instrum. Methods Phys. Res. A \textbf{922}, 98 (2019).

		\bibitem{McKeenHeating2021}
		D. McKeen, M. Pospelov, and N. Raj, Phys. Rev. Lett. \textbf{127}, 061805 (2021).

	\end{thebibliography}
\end{document}